\documentclass[trackchanges,twocolumn]{aastex701}

\usepackage{lineno}
\usepackage{amssymb}
\usepackage{amsmath}
\usepackage{booktabs}
\usepackage{hyperref}
\usepackage{multirow}
\usepackage{times}
\PassOptionsToPackage{hyphens}{url}\usepackage{hyperref}
\usepackage[hyphenbreaks]{breakurl}

\newcommand{\ngc}{NGC\,4051}
\newcommand{\xmm}{{\it XMM-Newton}}
\newcommand{\xrism}{{\it XRISM}}
\newcommand{\nustar}{{\it NuSTAR}}

\newcommand{\arcs}{\hbox{$^{\prime\prime}$}}

\newcommand{\ls}
{\mathrel{\hbox{\rlap{\hbox{\lower4pt\hbox{$\sim$}}}\hbox{$<$}}}}
\newcommand{\gs}
{\mathrel{\hbox{\rlap{\hbox{\lower4pt\hbox{$\sim$}}}\hbox{$>$}}}}

\begin{document}

\title{Winds of Change:  \xrism\ Resolve X-ray spectroscopy  of NGC 4051}
\shorttitle{X-ray spectroscopy of NGC 4051}

\author[]{James N.\ Reeves} 
\affiliation{Institute for Astrophysics and Computational Sciences, Department of Physics, The Catholic University of America, Washington, DC 20064, USA}
\affiliation{INAF, Osservatorio Astronomico di Brera, Via Bianchi 46, I-23807 Merate (LC), Italy}
\email{james.n.reeves456@gmail.com}

\author[]{Shoji Ogawa}
\affiliation{Institute of Space and Astronautical Science (ISAS), Japan Aerospace Exploration Agency (JAXA), Sagamihara, Kanagawa 252-5210, Japan
}
\email{ogawa.shohji@jaxa.jp}

\author[]{Tracey~J.~Turner}
\affiliation{Eureka Scientific, Inc., 2452 Delmer Street Suite 100, Oakland, CA 94602-3017, USA
}
\email{turnertjane@gmail.com}

\author[]{Valentina Braito}
\affiliation{INAF, Osservatorio Astronomico di Brera, Via Brera 20, I-20121 Milano, Italy}
\affiliation{Dipartimento di Fisica, Università di Trento, Via Sommarive 14, Trento 38123, Italy}
\affiliation{Institute for Astrophysics and Computational Sciences, Department of Physics, The Catholic University of America, Washington, DC 20064, USA}
\email{valentina.braito@gmail.com}

\author[]{Satoshi Yamada}
\affiliation{The Frontier Research Institute for Interdisciplinary Sciences, Tohoku University, Aramaki, Aoba-ku, Sendai, Miyagi 980-8578, Japan
}
\affiliation{Astronomical Institute, Tohoku University, Aramaki, Aoba-ku, Sendai, Miyagi 980-8578, Japan
}
\email{satoshi.yamada@terra.astr.tohoku.ac.jp}

\author[]{Steven B.\ Kraemer}
\affiliation{Institute for Astrophysics and Computational Sciences, Department of Physics, The Catholic University of America, Washington, DC 20064, USA}
\email{kraemer@cua.edu}

\author[]{Hirofumi Noda}
\affiliation{Astronomical Institute, Tohoku University, Aramaki, Aoba-ku, Sendai, Miyagi 980-8578, Japan
}
\email{hirofumi.noda@astr.tohoku.ac.jp}

\author[]{Anna Trindade Falc{\~a}o}
\affiliation{NASA Goddard Space Flight Center, Code 662, Greenbelt, MD 20771, USA}
\email{annatrindadefalcao@gmail.com}

\author[]{Martin Elvis}
\affiliation{Harvard-Smithsonian Center for Astrophysics, 60 Garden St., Cambridge, MA 02138, USA}
\email{melvis@cfa.harvard.edu}

\author[]{Giuseppina Fabbiano}
\affiliation{Harvard-Smithsonian Center for Astrophysics, 60 Garden St., Cambridge, MA 02138, USA}
\email{gfabbiano@cfa.harvard.edu}



\begin{abstract}

NGC\,4051 is a nearby (16.7\,Mpc), Narrow Line Seyfert 1 galaxy (NLS1), which has a low black hole mass of $10^6$\,M$_{\odot}$. It is also known for its rapid X-ray variability, on timescales of kilo-seconds and has a 
complex, multi component wind in both the soft X-ray and Fe K bands. Here we present the first high resolution \xrism\ Resolve spectrum of NGC\,4051, which was captured in a historically bright state for a 150\,ks 
exposure. \xrism\ resolves two blue-shifted Fe K shell absorption troughs in the mean spectrum, which can be ascribed to H-like iron and arises from 
two outflow components with outflow velocities of 0.025c and 0.04c. A time dependent spectral analysis shows that the iron K absorption is variable on timescales of less than a day, 
increasing in velocity over the duration of the observation. The velocity changes may be explained either by the passage of two separate transiting absorbers, of different velocities, or by a single accelerating outflow 
of approximately constant column density. 
In the latter case, the wind acceleration is likely to be too large to be caused by radiation pressure and instead magnetic driving is favored to accelerate the wind up to 0.04c. The outflow can originate from an accretion disk wind, whose kinetic power is sub-Eddington in contrast to recent examples of winds from powerful, luminous quasars observed by \xrism.  

\end{abstract}




\section{Introduction}

X-ray outflows are common in type I Active Galactic Nuclei (AGN). In Seyfert 1s, warm absorbers are observed in soft X-ray grating spectra, originating from low velocity outflows at typically hundreds to a few thousand kilometers per second \citep{Crenshaw03,Blustin05,Laha21}. In the iron K band, CCD resolution spectra often reveal absorption from K-shell lines of He and H-like iron \citep{Chartas02,Chartas03,Pounds03,Reeves03,Tombesi10,Gofford13,Nardini15,Matzeu23,Yamada24,Laurenti25}. 
This gas is of higher ionization than the warm absorbers and their velocities can reach 0.1c or higher in the so-called Ultra Fast Outflows (UFOs), indicating that they originate from matter ejected close to the super-massive black hole \citep{Tombesi13}. 
As a result of their enhanced energetics, UFOs can play an important role in black-hole and host galaxy co-evolution \citep{King03,King10,Hopkins16}. 

\begin{deluxetable*}{lccccc}
\tablecaption{Observation log of the 2025 NGC 4051 campaign}
\tablewidth{0pt}
\tablehead{
\colhead{Instrument} & \colhead{Obs. ID} & \colhead{Start Date} & \colhead{End Date} & \colhead{Exposure$^{a}$} & \colhead{Count Rate$^{b}$}}
\startdata
XRISM Resolve & 201066010 & 2025-05-16 18:01:30 & 2025-05-20 06:28:10 & 134 & $0.437\pm0.002$\\
XRISM Xtend & 201065010 & 2025-05-16 17:54:41& 2025-05-20 06:59:27 & 157 & $6.798\pm0.007$\\
NuSTAR FPMA$+$B & 61101003002 & 2025-05-15 16:24:51 & 2025-05-19 05:41:06 & 316 & $1.301\pm0.003$\\
XMM-Newton~1 EPIC-pn & 0960150201 & 2025-05-15 21:37:36 & 2025-05-17 05:53:32 & 80.5 & $34.10\pm0.02$\\
XMM-Newton~2 EPIC-pn & 0960150301 & 2025-05-17 16:39:34 & 2025-05-19 05:13:22 & 85 & $35.28\pm0.02$\\
\enddata
\tablenotetext{a}{Net exposure time in units of ks.}
\tablenotetext{b}{Observed background-subtracted count rate in units of ct\,s$^{-1}$. The \nustar\ rate is for FPMA$+$B combined.}
\label{tab:obs_log}
\end{deluxetable*}

The Resolve calorimeter onboard  {\it XRISM} provides a Full Width at Half Maximum (FWHM) spectral resolution of $\sim 4.5$\,eV \citep{Tashiro25}.
Resolve thus provides precise measurements of highly ionized outflows measured in the iron K band, as is seen during initial observations of AGN; e.g. PDS\,456 \citep{xrism25}, 
NGC\,3783 \citep{Mehdipour25} IRAS\,05189--2524 \citep{Noda25}, PG\,1211+143 \citep{Mizumoto25}, NGC\,4151 \citep{Xiang25}, NGC\,3516 \citep{Juranova25}, MCG\,$-$6$-$30$-$15 \citep{Brenneman25} and NGC\,1365 \citep{Z26}. 
These observations have revealed outflow velocities ranging from just a few hundred kilometers per second up to a substantial fraction of the speed of light, over a range of radial distances and 
often with complex and clumpy structures. 

Not only can Resolve precisely measure the outflow kinematics, it can also probe the outflow variability on timescales of a day or less, potentially providing insight into 
the wind acceleration mechanisms.  Notably, the \xrism\ observation of the Seyfert galaxy NGC\,3783 captured the emergence of an ultra fast outflow, 
which appears to accelerate from $0.05c$ to $0.30c$, a rate at which \cite{Gu25} infer to be too large to be due to radiation pressure, favoring magnetic hydrodynamical (MHD) driving mechanisms \citep{BP82,Fukumura10}.


The subject of this paper is NGC\,4051, a bright, nearby (16.6\,Mpc, $z=0.002336$; \citealt{Yuan21})  Narrow Line Seyfert\,1 AGN with 
bolometric luminosity  $\sim 10^{43.4} {\rm  erg\,  s^{-1}}$  \citep{Blustin05}.  
Measurements of the lag time between variations in the continuum
and H$\beta$ emission line lead to a measured  radius  for the optical broad line region (BLR) of $R_{\rm BLR} \approx 2$\,light-days 
and a black hole mass of $M_{\rm BH} \approx 10^{6}M_{\odot}$ \citep{Denney09,Fausnaugh17}, with typical 0.3~dex uncertainties.

 
NGC\,4051 has a strong Fe K$\alpha$ line and a well-studied multi-zoned X-ray absorber, originating from a variable nuclear wind  \citep{Pounds04,Krongold07,Sera25}. 
It also shows pronounced iron K shell absorption, as revealed by earlier CCD resolution spectra \citep{Lobban11,PV12}, which are blue-shifted by typically $5000-7000$\,km\,s$^{-1}$. 
A further faster UFO component can also be present in some observations, reaching velocities of up to $\sim$0.12c \citep{Tombesi10} and which appear to be variable on a $\sim2$~day timescale \cite{PV12}.   
The existence of these absorbers are supported by an analysis of the available Chandra HETG spectra on NGC\,4051 \citep{Ogorzalek22}, which show multiple absorber zones spanning the velocity range from a few hundred kilometers 
per second up to 0.1c.


\begin{figure*}
\begin{center}
\rotatebox{-90}{\includegraphics[width=6.5cm]{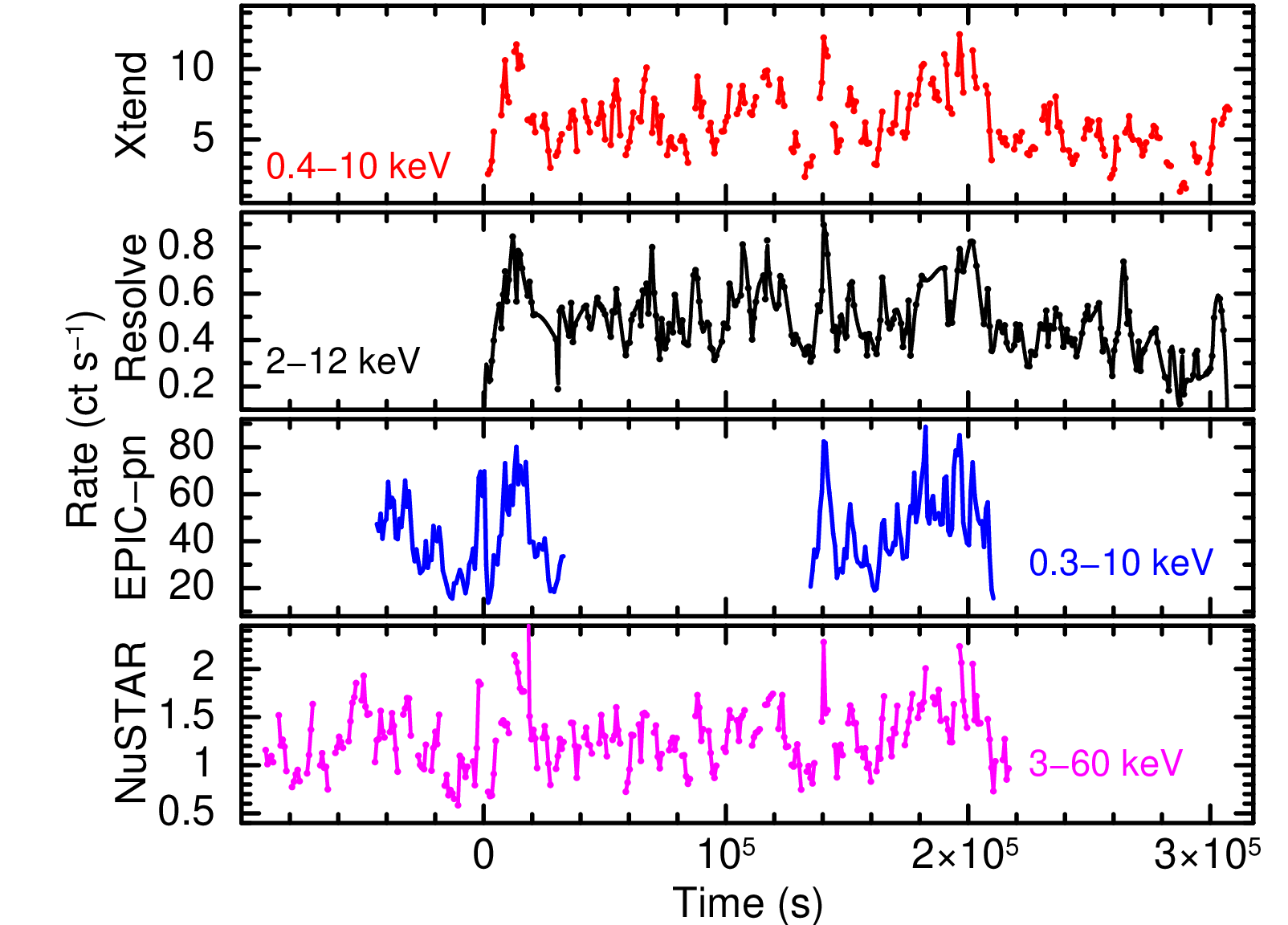}}
\rotatebox{-90}{\includegraphics[width=6.5cm]{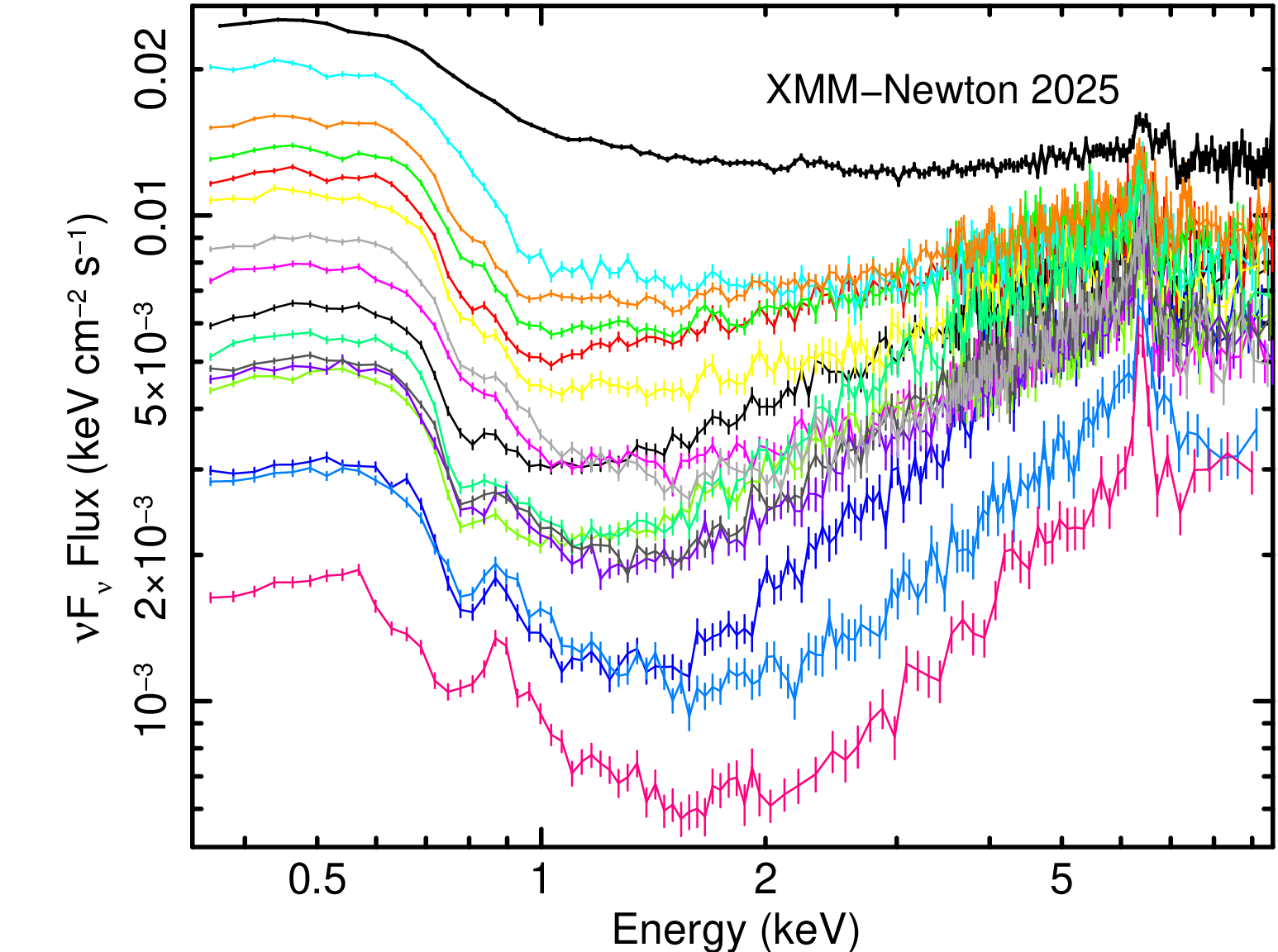}}
\rotatebox{-90}{\includegraphics[width=6.2cm]{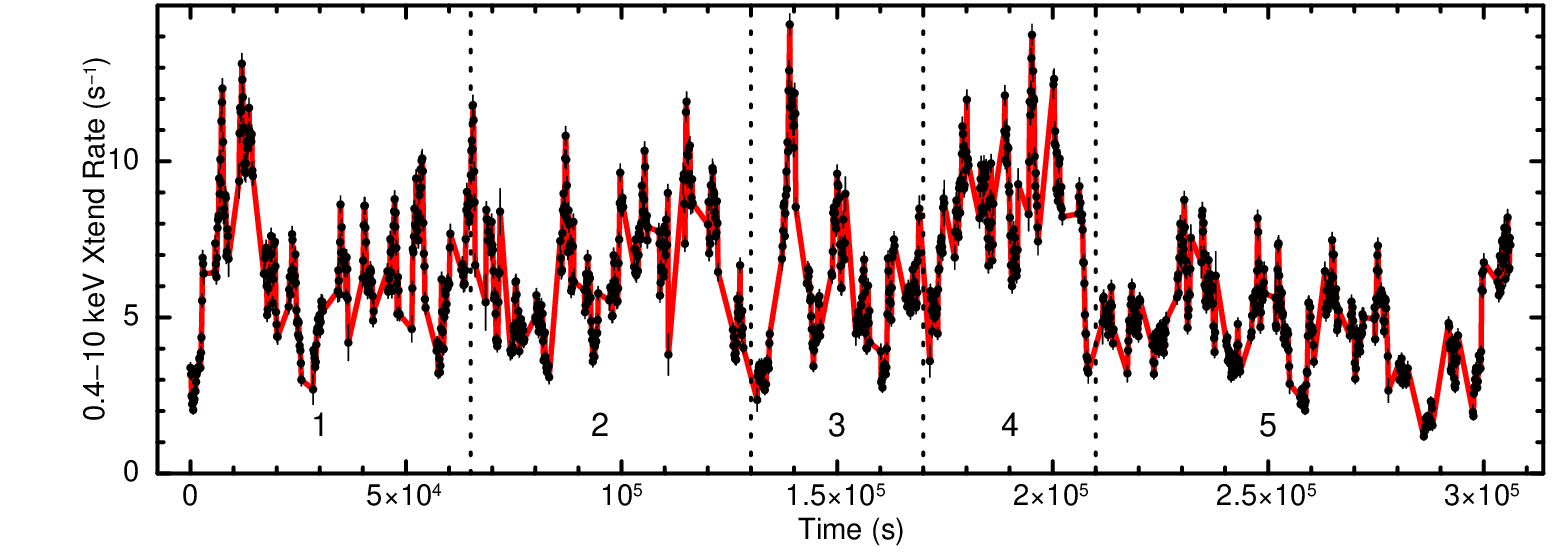}}
\end{center}
\caption{The X-ray variability of \ngc. The upper left panels show the 2025 light-curves (top to bottom) from Xtend, Resolve, EPIC-pn and NuSTAR. Time is relative to the start time of the \xrism\ Resolve exposure. The lower panel shows a zoom in of the 0.4-10\,keV Xtend lightcurve, with finer time binning of 128\,s per bin. Multiple short timescale kilo-second flares are present. The vertical dashed lines denote the time intervals used for the time sliced spectral analysis. The data points are shown in black and the bold red line connects the points.  The upper right plot highlights the long term spectral variability, where the 2025 \xmm\ EPIC-pn spectrum (bold black points) is observed at a historical high flux compared to the 15 EPIC-pn spectra taken in 2009.}
\label{fig:lightcurve}
\end{figure*}

A \xrism\ observation of NGC~4051 was conducted in May 2025, supported by  {\it NuSTAR} and {\it XMM-Newton}, as is summarized in Table 1 and illustrated in Figure~1 for each instrument. 
The AGN shows a high level of variability throughout the campaign. 
Here we present the signature and variability of the iron K band absorber at high resolution.   
Section~2 describes the observations. Section 3 presents the mean spectral analysis, providing a characterization of the absorption line profiles measured by Resolve. 
These appear to originate from mini-UFOs, with outflow velocities from $0.02-0.04$c. Section~4 
describes the time variable iron K absorption, where an increase in outflow velocity occurs during the \xrism\ exposure. Section 5 interprets this in terms of either transiting material or an accelerating outflow.  
Future work will present the detailed analysis of the iron K band emission profiles and the soft X-ray grating spectra.

\section{Data Analysis}

\subsection{XRISM}

A  {\it XRISM} observation was conducted for NGC 4051 with a total duration of $\sim 300$\,ks, using both the Resolve micro-calorimeter \citep{Kelley25} and the Xtend CCDs \cite{Noda25b}.
Data were screened to  exclude periods during Earth occultation and  passages through the 
 South Atlantic Anomaly
(SAA). In addition, times within 4300\,s of the initiation of the cryogenic cooler recycling, and visibility of the Earth's sunlit limb were also excluded.  
After event screening the effective exposure time for Resolve was 134\,ks and 157\,ks for Xtend. 
 
For the Resolve spectra and responses, only the Hp (High-resolution 
Primary) events were utilized and low resolution secondary events (${\rm L}_{\rm S}$ pseudo events) were excluded from the analysis.  
The data reduction used pipeline software version 03.01.014.010 with calibration database {\sc caldb} 20241115. 
In addition to the calibration pixel, pixel 27 is excluded from analysis as it  exhibits random
gain spikes of unknown origin. 
The on-board Fe--55 calibration pixel confirms an in-flight Resolve resolution of $4.55\pm 0.05$\,eV (FWHM), with an energy scale uncertainty of $\pm 1$\,eV.  

A stacked Non X-ray Background (NXB) spectrum\footnote{see https://heasarc.gsfc.nasa.gov/docs/xrism/analysis/nxb/index.html} 
was generated and found to be negligible compared to the AGN;  the total Resolve count-rate is $0.442\pm 0.002~{\rm ct\, s^{-1}}$ from 2--10\,keV, of which the NXB contribution is only 1\%. 
The mean Resolve spectrum was binned to 2\,eV per bin for spectral fitting to over-sample the FWHM resolution and was also re-sampled to 10\,eV and 20\,eV
per bin for plotting purposes. For the time sliced analysis, the coarser binning is used. 

An Xtend CCD spectrum was extracted from a circular
source region of radius 2.5' and background data from 
several boxes that sampled source-free regions of the detector. Xtend yielded a source count rate $1.295\pm 0.003~{\rm cts\, {\rm s}^{-1}}$ (2--10\,keV) or 
$6.798\pm 0.007~{\rm cts\, {\rm s}^{-1}}$ (0.4--10\,keV).  
 
 \subsection{NuSTAR}
 
 {\it NuSTAR} observed NGC\,4051 starting about 1 day before the \xrism\ exposures and overlapped both \xmm\ exposures; 
 see Table~1 and Figure~1. 
 {\it NuSTAR} Focal Plane Modules A  (FPMA) and
B  (FPMB) offer high signal spectral data from $3-60$\,keV for NGC 4051, facilitating 
determination of  the primary continuum form. 
This yielded count rates of $0.659\pm 0.002~{\rm cts\, {\rm s}^{-1}}$ from FPMA and $0.640\pm 0.002~{\rm cts\, {\rm s}^{-1}}$ from FPMB. 

\begin{figure*}
\begin{center}
\rotatebox{-90}{\includegraphics[width=5cm]{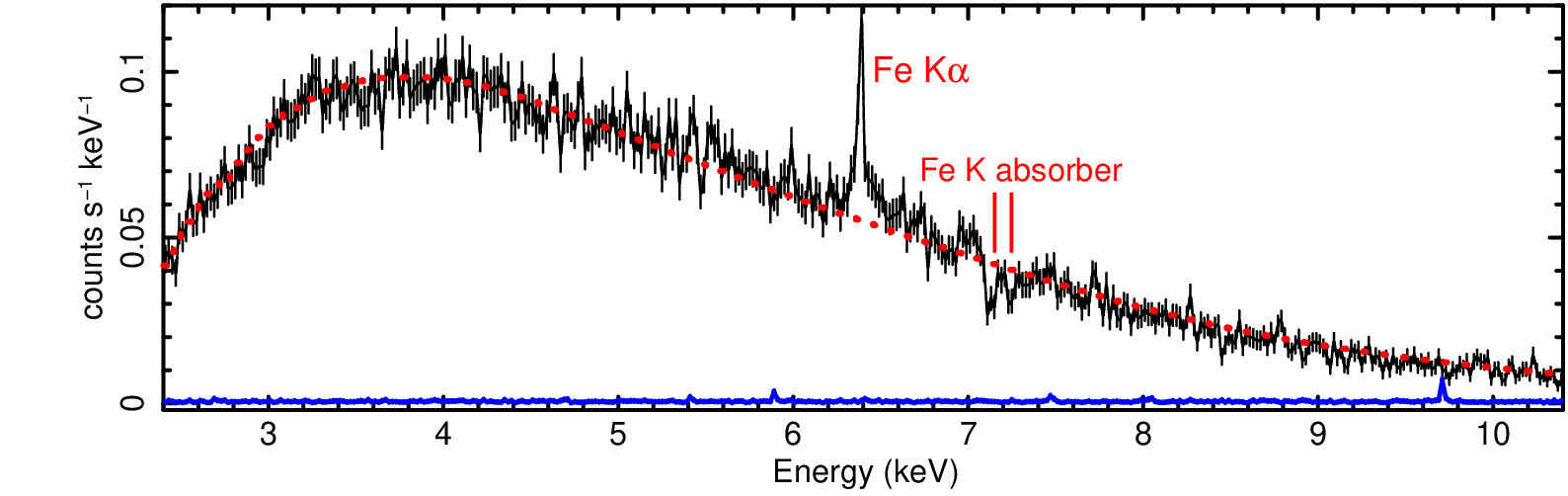}}
\rotatebox{-90}{\includegraphics[width=6.3cm]{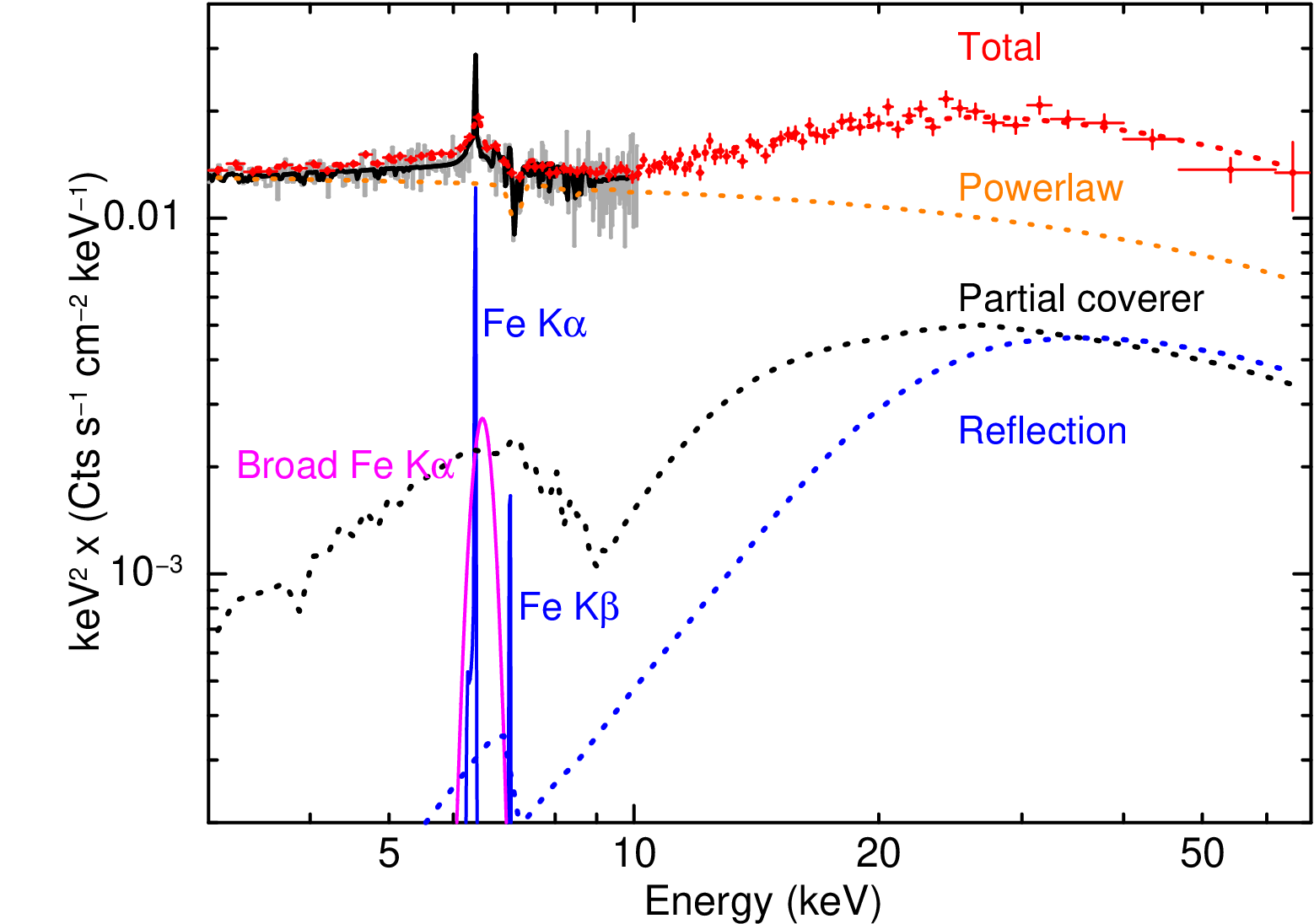}}
\rotatebox{-90}{\includegraphics[width=6.3cm]{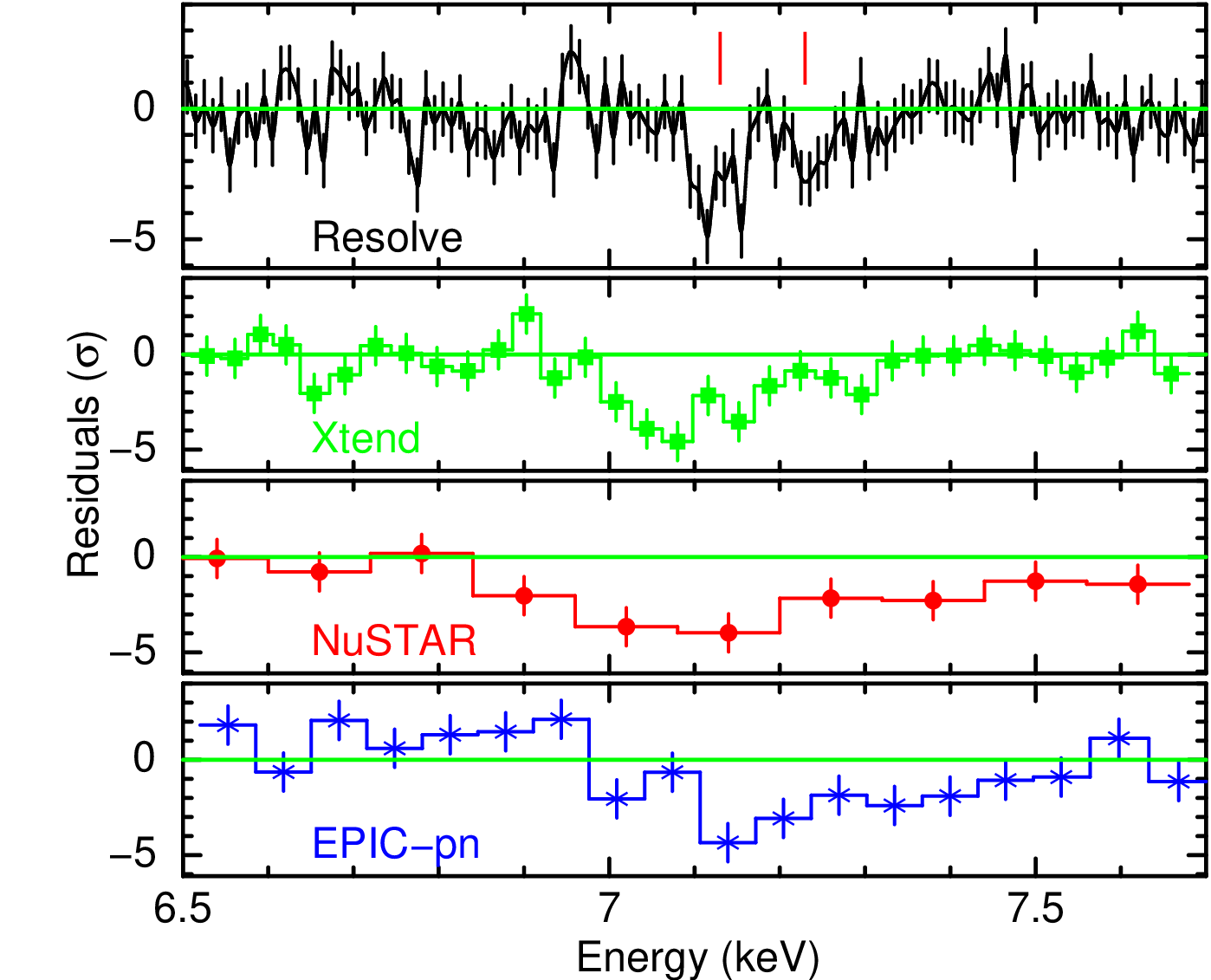}}
\end{center}
\caption{Time averaged spectra of NGC 4051 from the 2025 campaign. The top panel shows the \xrism\ Resolve spectrum (black), with the best-fit continuum model plotted as a red dotted line and the background (NXB) spectrum is shown in blue. The position of the strong Fe K$\alpha$ line and two absorption troughs are marked in red. The lower left plot shows the broad-band spectrum fitted with the model of Table~2. Here, Resolve is shown in black (grey errors) and \nustar\ FPMA in red. The individual model components (dotted lines) are as marked. The lower-right panels show the residuals of the data to the model (in $\sigma$ units), but without including any outflowing absorbers. The two absorption troughs are apparent in Resolve (red vertical markers) and appear as a single trough in the lower resolution spectra. 
Here, EPIC-pn (sequence 2) is also plotted in blue and \xrism\ Xtend in green.}
\label{fig:average}
\end{figure*}

\subsection{XMM-Newton}

\xmm\ observed NGC\,4051 over two satellite orbits, one centered on the start of the \xrism\ exposures and the second about half-way through (Figure~1). 
The spectra were extracted from circular regions, of radius 27\arcs\ for the source and 45\arcs\ for the background in Small Window mode. 
These were screened for periods of high background, resulting in net exposures of 80.5\,ks and 85\,ks for sequences 1 and 2 respectively. 
The corresponding net count rates are $34.10\pm0.02$\,cts\,s$^{-1}$ (sequence~1) and $35.28\pm0.02$\,cts\,s$^{-1}$ (sequence~2) from 0.3--10\,keV.

Figure 1 (upper right panel) compares the 2025 EPIC-pn spectrum, in $\nu{\rm F}_{\nu}$ flux, with the series of $15\times40$\,ks observations with \xmm, taken over a month in 2009 \citep{PV12}. 
In 2025, NGC\,4051 was caught in a high flux state, at the top end of the range of 2--10\,keV fluxes 
of $0.5-3.5\times10^{-11}$\,ergs\,cm$^{-2}$\,s$^{-1}$. At lower fluxes the spectra appear harder, possibly as a result of the increased contribution of reprocessed emission, via absorption and scattering \citep{Miller10}. In contrast, the 2025 observations capture a soft, bright, continuum dominated state. Note there are no strong changes in source hardness within the 2025 observations.

\section{The Time-Averaged Spectrum}

The $2-10$\,keV range of the Resolve and Xtend data were fit together with {\it NuSTAR} data from  $3-60$\,keV in order to first define the time-averaged hard X-ray spectrum of NGC\,4051.   
All models included the Galactic line-of-sight absorption, ${\rm N_{H, Gal}=} 1.19 \times 10^{20}{\rm cm^{-2}}$ \citep{Dickey90}, via the \textsc{xspec} \textsc{tbabs} model \citep{Wilms00}. 
The data were fit with the C-statistic \citep{Cash79}.
The spectra were binned to 100 counts per bin, except for Resolve as noted in Section~2.
Errors are quoted at 90\% confidence for one parameter of interest ($\Delta C = 2.71$) and all parameters are in the rest-frame at $z=0.002336$.
For the luminosity, a distance of $D=16.6$\,Mpc is adopted, using the recent Cepheid variable estimate from \citet{Yuan21}. 

The Resolve spectrum, which captures a high flux state of NGC\,4051, is shown in the upper panel of Figure~2 and shows clear deviations from a powerlaw continuum in the Fe K-shell band. 
Fe K$\alpha$ emission is readily apparent, including both a narrow core of the line and velocity broadened wings. A detailed description of the Fe K emission 
modeling will be presented in paper~II.  

Two absorption troughs appear to be present at energies just above 7.1\,keV. These can be modeled by Gaussian absorption profiles at centroid energies of $7.14\pm0.01$\,keV and 
$7.26\pm0.02$\,keV.  The energy separation of the lines is too wide to be accounted 
for by the H-like Fe\,\textsc{xxvi} Ly$\alpha$ doublet at 6.952\,keV and 6.973\,keV (i.e. $\Delta E=0.021$\,keV) and it is too narrow to be explained by a combination of the H-like lines and the 
He-like Fe\,\textsc{xxv} resonance line at 6.700\,keV. This implies that the separation of the troughs derives from two different outflow velocity components of the absorber, e.g. 
each trough may arise from a velocity broadened blend of the two H-like doublet lines from each absorber. 
Note the contribution of the neutral iron K edge at 7.112\,keV is negligible, with an optical depth of $\tau<0.02$. 

The two troughs are broadened with a width of $\sigma=25^{+8}_{-4}$\,eV, which corresponds to $\sigma=1050^{+340}_{-170}$\,km\,s$^{-1}$ at 7.14\,keV. 
Note the widths of the two troughs were consistent with each other within the errors and were kept tied in the fits. 
Overall, the addition of the two troughs improves the fit statistic by $\Delta C=-35.8$ and $\Delta C=-16.6$ with respect to the Resolve data for the lower and higher energy components. 
According to the Akaike Information Criteria (AIC, \citealt{Akaike74,Tan12}), the troughs are significant with null hypothesis probabilities of $P_{\rm AIC}=3.4\times10^{-7}$ and $P_{\rm AIC}=1.8\times10^{-3}$ respectively. 
\footnote{The AIC is equivalent to ${\rm AIC} = 2k + {\rm C\,stat}$, where $k$ is the number of fitted parameters. The null hypothesis probability is found by the likelihood ratio test as $N = e^{\Delta {\rm AIC} / 2}$.}

\subsection{Photoionization Modeling} 
 
In order to model the absorption troughs and the broad-band spectrum between 2--60\,keV, 
photoionization models were constructed using \textsc{xstar} \citep{Kallman04} and the input SED of NGC\,4051 (see Appendix~A). 
The input continuum is approximated by a photon index slightly steeper than $\Gamma=2$, modified by an exponential cut-off (\textsc{cutoffpow}) at hard X-rays. 
The absorption tables were calculated for Solar abundances \citep{Asplund21} and turbulence velocities covering the range from $v_{\rm turb}=100-2700$\,km\,s$^{-1}$. 
The \textsc{mytorus} model \citep{Murphy09} accounts for the narrow Fe~K$\alpha_{1,2}$ and K$\beta$ emission lines, using the latest line tables as described in \citet{Yaqoob24} which incorporate 
the laboratory line profiles of \citet{H97}. These were then convolved with a Gaussian broadening function to describe their velocity widths. The much broader Fe K$\alpha$ wings are modeled by a separate Gaussian profile.
A scattered \textsc{mytorus} component models the 
associated Compton hump.

\begin{deluxetable}{lc}
\tabletypesize{\small}
\tablecaption{Spectral parameters for the mean spectrum.}
\tablewidth{0pt}
\tablehead{
\colhead{Parameter} & \colhead{Value}}
\startdata
\multicolumn{2}{c}{Cut-off Powerlaw}\\
$\Gamma$ & $2.05\pm0.03$\\
Normalization$^a$ & $2.11\pm0.13\times 10^{-2}$\\
$E_{\rm cut}$ & $110^{+95}_{-40}$\,keV \\
\hline
\multicolumn{2}{c}{MyTorus Line + Scattered}\\
Normalization$^a$ &  $0.91\pm0.18\times10^{-2}$\\
Column Density, $N_{\rm H}$ & $>6.3\times 10^{24}$\,cm$^{-2}$\\
Convolution, $\sigma_{\rm gauss}$$^b$ & $2.9^{+2.2}_{-1.8}$\,eV\\
FWHM width & $320^{+250}_{-200}$\,km\,s$^{-1}$\\
\hline
\multicolumn{2}{c}{Broad Iron K Line}\\
Broad Fe K Gaussian:-\\
Energy & $6.445\pm0.025$\,keV\\
Width, $\sigma_{\rm gauss}$ & $0.101^{+0.046}_{-0.030}$\,keV\\
FWHM width & $11000^{+5000}_{-3300}$\,km\,s$^{-1}$\\
Line Flux &  $1.9\pm0.4\times 10^{-5}$\,photons\,cm$^{-2}$\,s$^{-1}$\\
Equivalent Width (EW) & $52\pm11$\,eV\\
\hline
\multicolumn{2}{c}{Partial Covering Absorber}\\
Column Density, $N_{\rm H}$ & $1.14\pm0.11\times 10^{24}$\,cm$^{-2}$\\
Ionization, $\log\xi$$^c$ & $2.80\pm0.06$\\
Covering Fraction, $f_{\rm cov}$ & $0.37\pm0.03$\\
\hline
\multicolumn{2}{c}{Outflow Zone 1}\\
Column Density, $N_{\rm H}$ & $1.7\pm0.5\times 10^{23}$\,cm$^{-2}$\\
Ionization, $\log\xi$$^c$ & $4.8^{+0.2}_{-0.3}$$^t$\\
Turbulence velocity, $v_{\rm turb}$ & $1500^{+800}_{-600}$\,km\,s$^{-1}$$^t$\\
Outflow velocity, $v/c$  & $-0.0248\pm0.0014$\\ 
& ($-7400\pm400$\,km\,s$^{-1}$)\\
$\Delta C/\Delta \nu$$^d$ & $-57.6/4$\\ 
Null probability, $P_{\rm AIC}$$^e$ & $1.7\times10^{-11}$\\
\hline
\multicolumn{2}{c}{Outflow Zone 2}\\
Column Density, $N_{\rm H}$ & $6.8\pm3.7\times 10^{22}$\,cm$^{-2}$\\
Outflow velocity, $v/c$  & $-0.040\pm0.002$\\ 
& ($-12000\pm600$\,km\,s$^{-1}$)\\
$\Delta C/\Delta \nu$$^d$ & $-34.8/2$\\ 
Null probability, $P_{\rm AIC}$$^e$ & $2.0\times10^{-8}$\\
\hline
\multicolumn{2}{c}{Other Parameters}\\
C$^f$ (NuSTAR/Resolve) & $1.05\pm0.01$\\
C$^f$ (Xtend/Resolve) & $0.94\pm0.01$\\
Flux (2-10 keV) & $3.5\times10^{-11}$\,ergs\,cm$^{-2}$\,s$^{-1}$\\
\enddata
\tablenotetext{a}{Power-law flux in units of photons\,cm$^{-2}$\,s$^{-1}$\,keV$^{-1}$ at 1\,keV.}
\tablenotetext{b}{Velocity width at 6.4\,keV through Gaussian convolution.}
\tablenotetext{c}{Units of Ionization ($\xi$) are ergs\,cm\,s$^{-1}$. }
\tablenotetext{d}{Improvement in $\Delta C$ upon adding the component to the model. $\Delta\nu$ is the change in degrees of freedom.}
\tablenotetext{e}{Null hypothesis probability derived from the AIC statistic.}
\tablenotetext{f}{Multiplicative cross normalization factor.}
\tablenotetext{t}{The ionization and turbulence are tied between zones 1 and 2.}
\label{tab:mean}
\end{deluxetable}

The phenomenological form of the model is simply:-\\

\noindent $\textsc{constant} \times \textsc{tbabs} \times [(2\,\textsc{xstar}_{\rm Fe} \times \textsc{zxipcf} \times \textsc{cutoffpow}) + \textsc{mytorus}_{\rm scatt} + \textsc{mytorus}_{\rm line}+ \textsc{gauss}].$\\

\noindent Here the cut-off power-law continuum is absorbed by two outflowing photoionized absorbers ($\textsc{xstar}_{\rm Fe}$). A  lower ionization partial covering absorber, represented by the \textsc{zxipcf} model in \textsc{xspec} \citep{Reeves08}, is also included, whereby a fraction ($f_{\rm cov}$) of the continuum is intercepted by the absorber and the remainder $(1-f_{\rm cov})$ is unattenuated. 
A constant cross normalization factor between instruments is included in front of all of the components. 

The spectral parameters are listed in Table~2 and the fit to the data is plotted in Figure~2 (lower left). The overall fit statistic is $C/\nu=5239.1/5184$. 
The X-ray power-law has a photon index of $\Gamma=2.05\pm0.03$, with a high energy cut-off of $E_{\rm cut}=110^{+95}_{-40}$\,keV which is typical of Seyfert galaxies observed 
with \nustar\ \citep{Middei19}.
The two \textsc{xstar} absorbers are both outflowing, 
with velocities of $v/c=-0.0248\pm0.0014$ and $v/c=-0.040\pm0.002$ which corresponds to $v_{\rm out1}=-7400\pm400$\,km\,s$^{-1}$ and $v_{\rm out2}=-12000\pm600$\,km\,s$^{-1}$. 
These velocities lie near to the boundary of the definition of an ultra fast outflow at $-10000$\,km\,s$^{-1}$ \citep{Tombesi10}. No other Fe K absorption components are required. 

The absorbers are highly ionized, with an ionization parameter of $\log\xi=4.8^{+0.2}_{-0.3}$\,erg\,cm\,s$^{-1}$. Indeed, negligable He-like absorption is apparent and the absorption arises mainly from 
H-like Fe. Neither absorber imparts any significant opacity in the lower energy part of the Resolve band between 
2--4\,keV, nor is any predicted in the soft X-ray band below 2\,keV. 
For simplicity, we have assumed both outflowing zones have the same ionization parameter, which are otherwise consistent within errors.
The high ionization parameter, along with the column densities that are observed ($N_{\rm H}\approx10^{23}$\,cm$^{-2}$) are typical of UFOs in nearby AGN \citep{Tombesi10,Gofford13}.
The turbulence velocity was found to be $v_{\rm turb}=1500^{+800}_{-600}$\,km\,s$^{-1}$ (equivalent to $\sigma_{\rm v}=1100^{+600}_{-400}$\,km\,s$^{-1}$). At this velocity width, the individual Fe\,\textsc{xxvi} components are not separated into their constituent doublet lines and each component appears as a single, velocity broadened H-like trough.

Against the full dataset, each of the outflowing zones is highly significant with $\Delta C/\Delta\nu=-57.6/4$ (slower zone) and $\Delta C/\Delta\nu=-34.8/2$ (faster zone). 
These correspond to null hypothesis probabilities of $1.7\times 10^{-11}$ and $2.0\times10^{-8}$ respectively, according to the AIC. The absorbers are also independently detected across all instruments; e.g. see Figure~2 (lower right panel), which shows the residuals of the individual Resolve, Xtend and \nustar\ and \xmm\ spectra. 
However, only Resolve separates the absorption into two kinematically distinct components, which appear as a single trough in the CCD resolution data. 

Partial covering absorption has been claimed previously in \ngc, in particular to account for pronounced spectral curvature observed in lower flux observations of this AGN which display much harder, more absorbed spectra \citep{Terashima09,Miller10,Lobban11}. 
In this high flux observation, the effects of the partial coverer upon the spectrum are modest. While its column density is high ($N_{\rm H}\approx10^{24}$\,cm$^{-2}$, Table~2), its line of sight covering fraction is very low 
($f_{\rm cov}=0.37\pm0.03$). As a result of its high column yet low covering, its transmission between 2--10\,keV 
is $<10$\% of the direct power-law continuum (see Figure~2, left). Subsequently it imparts no detectable Fe K absorption lines nor edges upon the Resolve spectrum. Its primary effect is in modeling an additional hard X-ray excess seen in the {\it NuSTAR} spectrum above 10\,keV, above that modeled by the \textsc{mytorus} reprocessor. 
The physical origin of this hard excess component will be deferred to future works. 


\section{Time Sliced Spectral Analysis}

The Resolve spectrum was then sliced into five time intervals in order to characterize the variability of the outflowing absorber. These are slice 1 ($0-65$\,ks), slice 2 ($65-130$\,ks), 
slice 3 ($130-170$\,ks), slice 4 ($170-210$\,ks) and slice 5 ($210-310$\,ks), as marked in Figure~1 (see lower panel). The sampling of the slices ensures there are at least $10^4$ counts in each Resolve spectrum. 
Here, slice~4 was chosen to cover the high portion of the 
X-ray lightcurve, with persistent X-ray flares and which is devoid of dips, while slice 5 covers a prolonged quiescent period following this. 
The time sliced Resolve spectra are plotted in Figure~3, zoomed into the 6--8\,keV band and are also compared to the mean spectrum.

\begin{figure*}
\begin{center}
\rotatebox{-90}{\includegraphics[height=11cm]{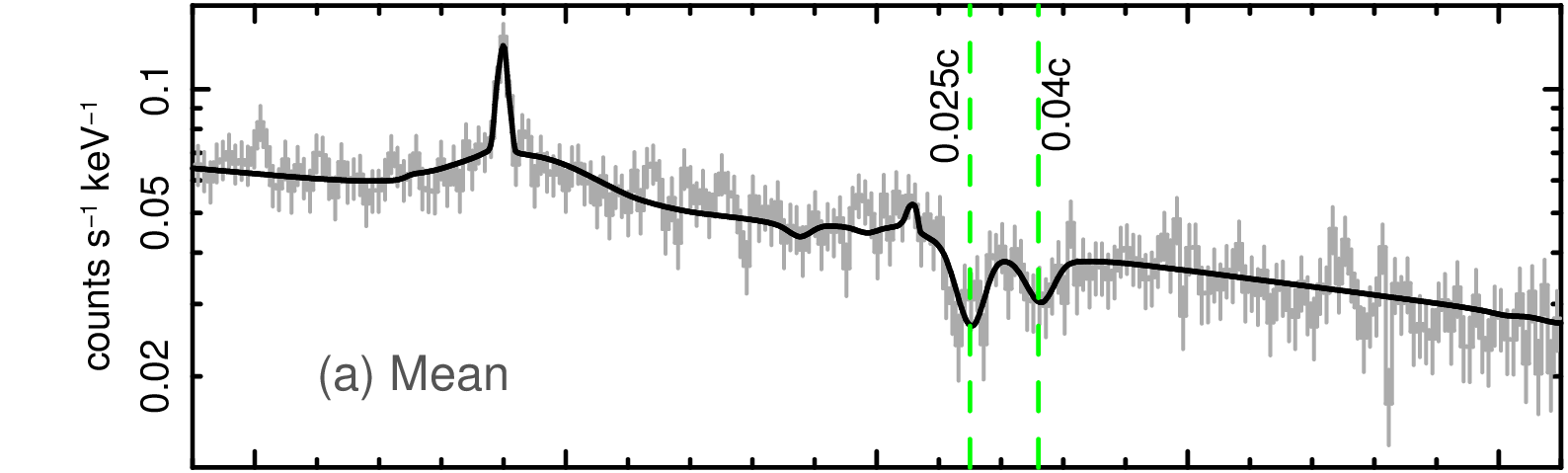}}
\rotatebox{-90}{\includegraphics[height=11cm]{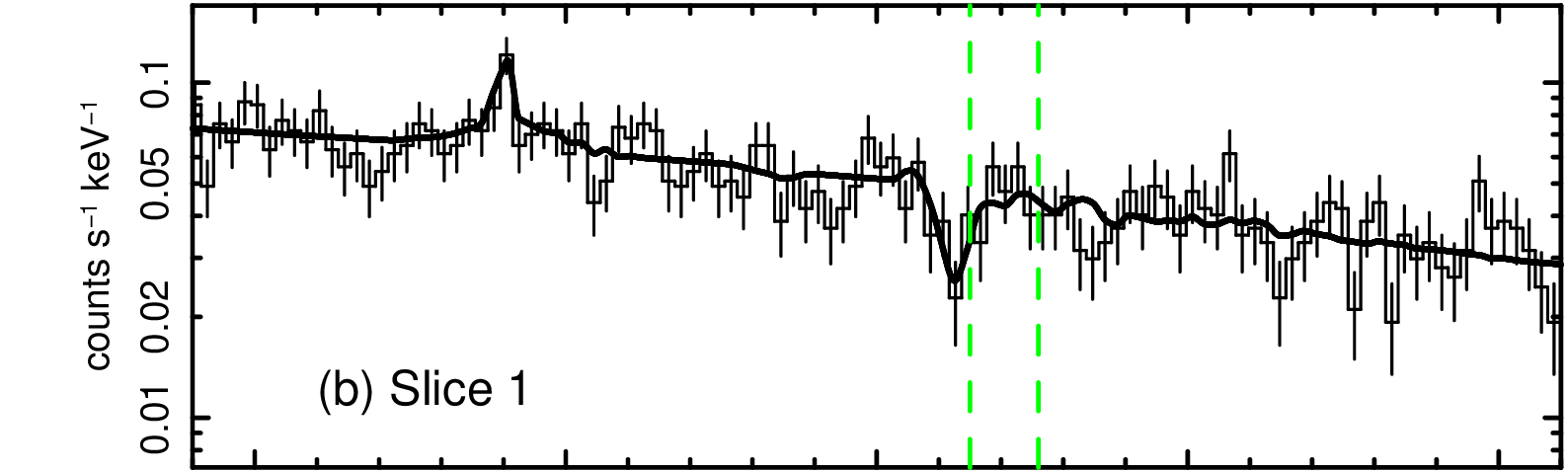}}
\rotatebox{-90}{\includegraphics[height=11cm]{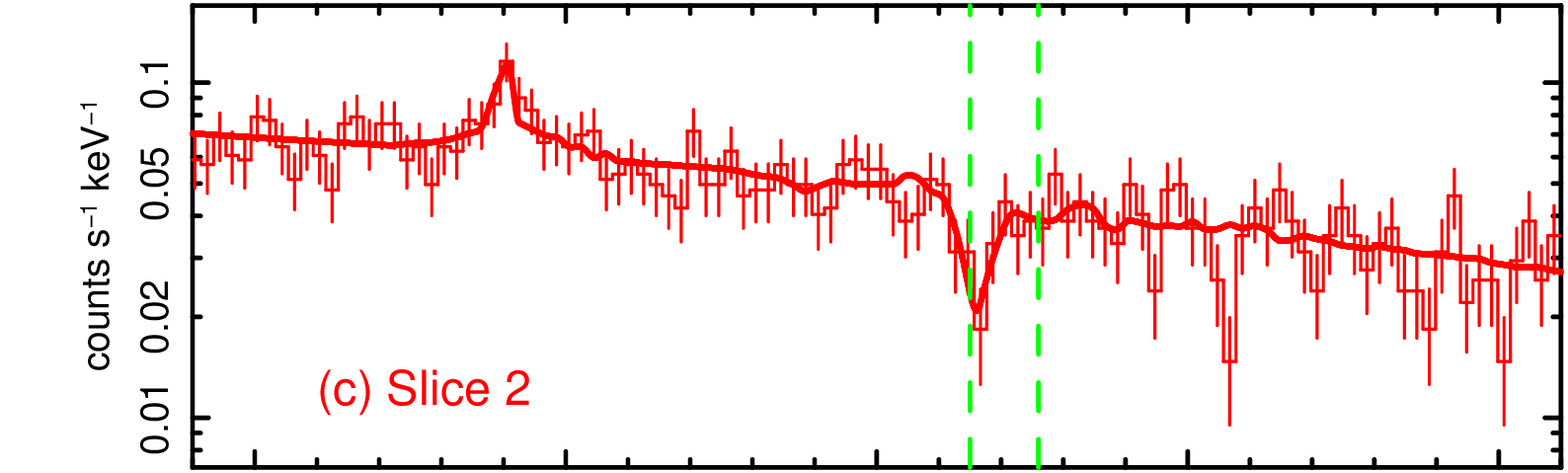}}
\rotatebox{-90}{\includegraphics[height=11cm]{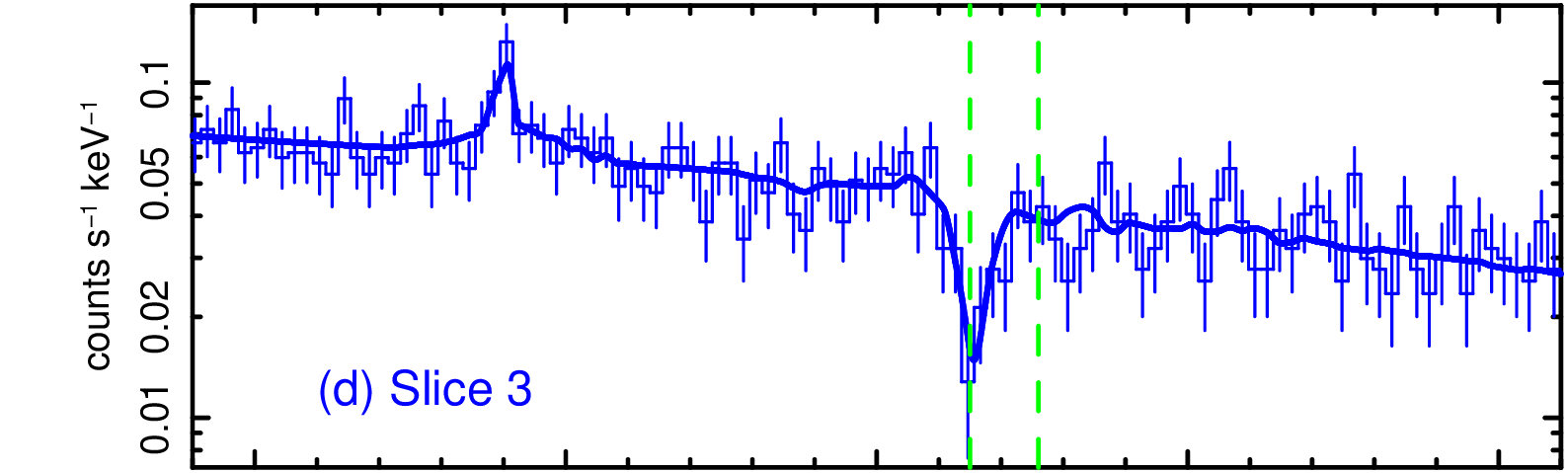}}
\rotatebox{-90}{\includegraphics[height=11cm]{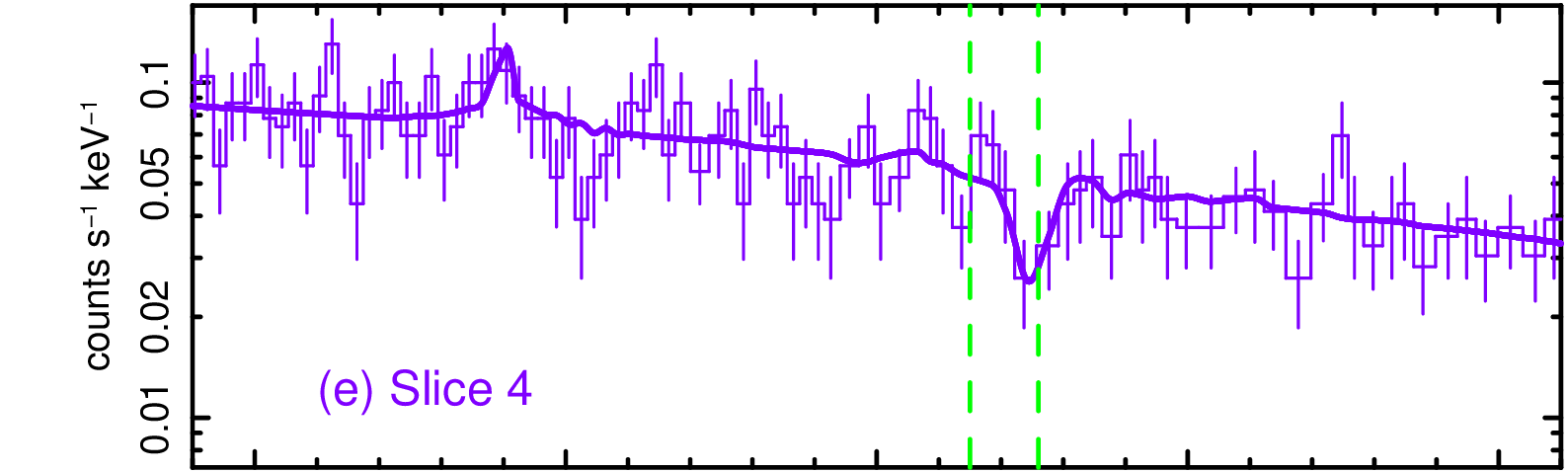}}
\rotatebox{-90}{\includegraphics[height=11cm]{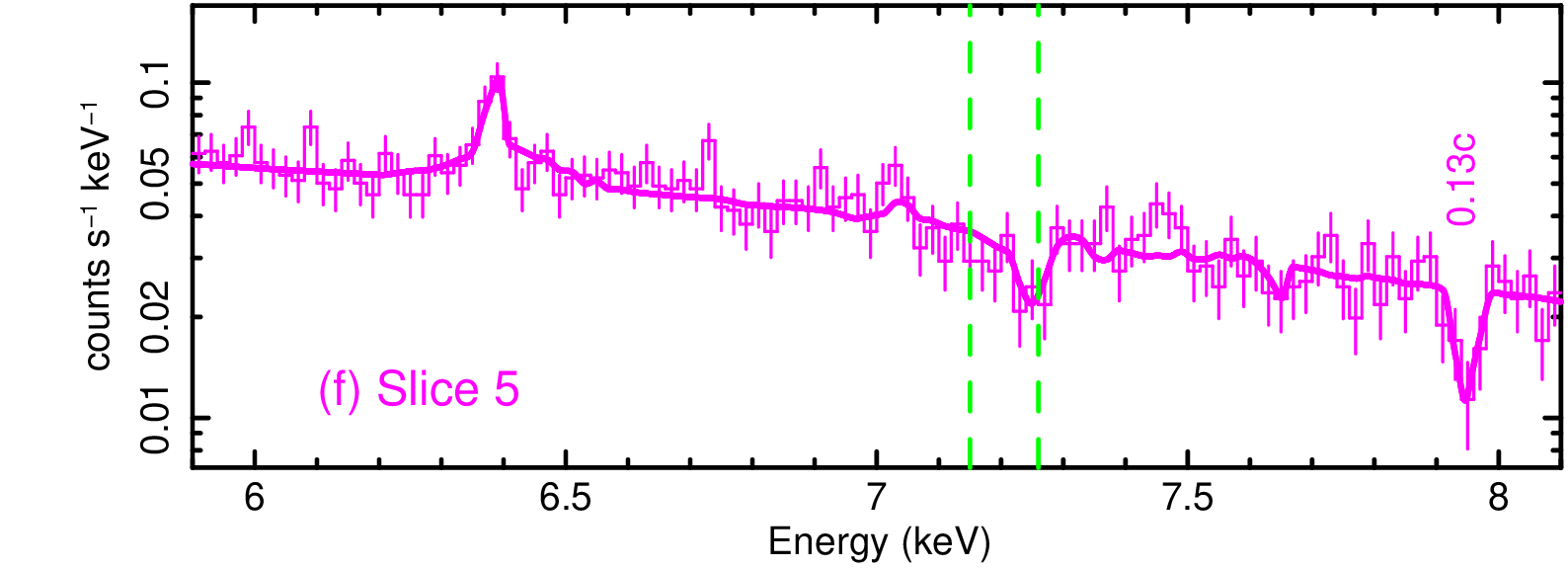}}
\end{center}
\caption{Time-dependent analysis of the Resolve spectrum, showing the mean spectrum (panel a) and the 5 slices (panels b--f) against the best fit absorber model as described in Table~3. The vertical dashed green lines mark the centroid energies of the two Fe K absorption troughs as measured in the mean spectrum at 7.14\,keV and 7.26\,keV respectively. The energy shift of the absorption lines are apparent between slices. In slices 1--3 the centroid energy of the absorption trough is consistent with the low 
energy trough at 7.14 keV, while in slices 4 and 5 the absorption line appears more blue-shifted coincident with the 7.26\,keV trough -- see Table~3 for values. This may suggest an increase in velocity in the Fe K absorber during the observation. Note slice 5 also shows a higher absorption trough at 7.96\,keV, which might originate from an ultra fast outflow at a velocity of $-0.13c.$}
\end{figure*}

To fit the slices, the same time-averaged model as listed in Table~2 was adopted. 
Asides from the continuum flux, all the parameters were fixed to the best fit values of the mean spectrum as no significant spectral variability of the continuum was found. For example, varying the photon index between the time sliced spectra did not result in any significant changes within the uncertainties; e.g. the lowest value was found in slice 3 ($\Gamma=2.01\pm0.05$) and the highest value in slice 4 
($\Gamma=2.11\pm0.06$), which are consistent within errors. Likewise, no change in the intensity of the narrow iron K$\alpha$ emission was found.

\subsection{Gaussian Line Search}

The variability of the absorption troughs was initially assessed by using single Gaussian trough in order to perform a blind trial search. This was performed for each slice, by varying the absorption line energy in 5\,eV energy increments (i.e. at the resolution of Resolve) over the 6.6--9.0\,keV band. 
The line widths were fixed to $\sigma=20$\,eV, consistent with the measured width in the mean spectrum, although allowing the widths to vary did not alter the search parameters.  
The scan results are shown in Appendix~B.

The resulting best-fit absorption line values from each slice are noted in Table~3. 
A minimum significance threshold of 95\% from the AIC was applied, which is equivalent to $\Delta C=-10$ for $\Delta\nu=2$. 
In addition, a set of 1000 Monte Carlo trials were also run to provide an independent estimate of the detection threshold. Here, the 95\% confidence level was found to occur at $\Delta C=-9.89$, consistent with the AIC. 
This is further described in Appendix~C.
The line subsequent search revealed a single, statistically significant trough in each of the five slices, which occur over the same expected energy range of 7.1--7.3\,keV as per the mean spectrum. 
The only exception is in slice 5, where an additional higher energy trough was also found at $>99$\% significance near to 8 keV. 
No other troughs were found to be significant at the $>95$\% confidence level. 

However, the centroid energies of the troughs are variable. In the first three slices, a single absorption trough was found from $7.12-7.16$\,keV, coincident with the first absorption trough in the mean spectrum, while the second higher energy absorption trough was not found. In contrast in the last two slices 4 and 5, the centroid energy of the absorption trough shifts to $7.25-7.26$\,keV, 
coincident with the second absorption line in the mean spectrum and no lower energy trough is found; see Figure 3 for a comparison.
Thus the sliced spectra reveal an increase in absorption 
line energy with time, with a significant jump appearing between slices 3 and 4. 
Superposition of all the individual troughs then result in the double trough seen in the average spectrum. 


The two EPIC-pn exposures confirm the absorption lines. \xmm\ sequence~1 covers the period from $-45$\,ks to 33\,ks (where $T=0$ corresponds to the start of the Resolve exposure), coinciding with slice 1. Sequence~2 corresponds to the period from $130-210$\,ks, coinciding with slice 3 and 4 obtained with Resolve. While EPIC-pn cannot resolve the profile, there is some indication of an energy shift from $7.11\pm0.05$\,keV to $7.22\pm0.04$\,keV; see Table~3 for details. 
The behavior is consistent with the increase in the absorption trough energy as seen in Resolve. The same behavior is not confirmed in the Xtend spectra, although they are consistent within the larger uncertainties.

\begin{deluxetable*}{lccccccccc}
\tabletypesize{\footnotesize}
\tablecaption{Summary of absorption zones fitted to the time sliced spectra.}
\tablewidth{0pt}
\tablehead{
\colhead{ } & \colhead{ } & \multicolumn{4}{c}{Gaussian fit} & \multicolumn{4}{c}{XSTAR fit}  \\
\cline{3-6} \cline{7-10} 
\colhead{Slice}  &  \colhead{$L_{\rm {2-10\; keV}}$} & \colhead{$E^{a}$} & \colhead{$EW$ } & \colhead{$\Delta C$ } & \colhead{$P_{\rm {AIC}}^b$ } & \colhead{log $N_{\rm H}$}  & \colhead{v/c } &\colhead{$\Delta C$ } &  \colhead{$P_{\rm {AIC}}^b$ } \\
\colhead{} &  \colhead{ $10^{41}$ erg s$^{-1}$}&   \colhead{keV} & \colhead{eV } & \colhead{ } & & \colhead{$10^{22}$\,cm$^{-2}$}  & \colhead{ } &\colhead{} &  \colhead{} 
}
\startdata						
Resolve 1	&12.7 &	$7.122^{+0.016}_{-0.014}$	&	$-24^{+10}_{-9}$	&	$12	$	&	$ 1.83\times 10^{-2}$	&	$	23.32^{+0.17}_{-0.24}$	&	$	-0.0222\pm0.0021$	&	$	14.6$	&	$4.99\times 10^{-3}$\\
Resolve 2	&12.3	 &     $7.162^{+0.011}_{-0.013}$	&	$-28^{+10}_{-9}	$	&	$15.5$	&	$ 3.18\times 10^{-3}$	&	$	23.41^{+0.15}_{-0.21}$	&	$	-0.0271\pm0.0015$	&	$	17.8$	&	$1.01\times 10^{-3}$	\\
Resolve 3	&12.1 &	$7.153\pm0.009	       $	&	$-37^{+9}_{-7}$		&	$26.4$	&	$ 1.37\times 10^{-5}$	&	$	23.58^{+0.14}_{-0.18}$	&	$	-0.0264\pm0.0015$	&	$	27.2$	&	$	9.17\times 10^{-6}$	\\
Resolve 4	&14.8 &	$7.252^{+0.013}_{-0.017}$	&	$-34^{+13}_{-10}$	&	$13.1$	&	$ 1.06\times 10^{-2}$	&	$	23.48^{+0.19}_{-0.30}$	&	$	-0.0388\pm0.0023$	&	$	11.6$	&	$	2.24\times 10^{-2}$	\\
Resolve 5	&9.9 &	$7.257^{+0.014}_{-0.018}$	&	$-24^{+12}_{-6}$		&	$13.1$	&	$ 1.06\times 10^{-2}$	&	$	23.20^{+0.16}_{-0.30}$	&	$	-0.0405\pm0.0022$	&	$	12.2$	&	$	1.66\times 10^{-2}$	\\
	&	  &	$7.963^{+0.014}_{-0.009}$	&	$-27^{+9}_{-10}$	&	$14.8$	&	$ 4.52\times 10^{-3}$	&	$	23.32^{+0.20}_{-0.25}$	&	$	-0.1336\pm0.0008$	&	$	17.2	$	&	$1.36\times 10^{-3}$	\\ 
\hline
EPIC-pn Seq~1 & & $7.11\pm0.05$ & $-22\pm9$ & $13.7$ & $7.8\times10^{-3}$ & $23.26^{+0.15}_{-0.22}$ & $-0.020\pm0.008$ & 18.1 & $8.6\times10^{-4}$ \\
EPIC-pn Seq~2 & & $7.22\pm0.04$ & $-31\pm9$ & $28.3$ & $5.3\times10^{-6}$ & $23.41^{+0.13}_{-0.15}$ & $-0.035\pm0.006$ & 35.1 & $1.8\times10^{-7}$ \\
\hline
Mean Resolve& &	$7.141^{+0.011}_{-0.009}$	&	$-26\pm6$		&	$35.8$	&	$3.38\times 10^{-7}$	&	$23.23^{+0.11}_{-0.15}$	&	$	-0.0255\pm0.0014$	&	$	37$	&	$	1.86\times 10^{-7}$\\
	  & 	&	$7.263^{+0.007}_{-0.025}$	&	$-17^{+6}_{-7}$		&	$16.6$	&	$1.84\times 10^{-3}$	&	$23.00^{+0.18}_{-0.22}$	&	$	-0.0398\pm0.0017$	&	$	15.9$	&	$	2.61\times 10^{-3}$\\
	 	\enddata
\tablenotetext{a}{For the Gaussian fits to the individual slices  we adopted a line width of $\sigma=20$\;eV. In the mean spectrum the width of the two absorption lines are tied, we measured $\sigma=25^{+8}_{-4} $\;eV.}
\tablenotetext{b}{Null probability of adding either the Gaussian absorption line or the \textsc{xstar} absorber, calculated via the AIC statistic.}
\label{tab:slices_gaus_xstar}
\end{deluxetable*}

\subsection{Modeling with an accelerating absorber}

Informed by the above results, a single \textsc{xstar} absorber was added to the baseline spectral model to probe its variations. 
The absorber outflow velocity was allowed to vary, along with its column, ionization and turbulence velocity across the five slices. 
However the latter two parameters were found to be consistent within the errors and were subsequently tied to the best-fit values obtained from fitting all five slices simultaneously. These are $\log\xi=4.8^{+0.2}_{-0.3}$ for the ionization and 
$v_{\rm turb}=1340^{+660}_{-600}$\,km\,s$^{-1}$ for the turbulence. 
Thus only the column density, outflow velocity and 
the power-law normalization were allowed to vary between the slices. 


A good fit to all five slices can be obtained with this simple prescription for the variability, 
with $C/\nu=1798.2/1809$ and the results are shown in Table~3. 
The absorber variability is driven by changes in the absorber velocity, which approximately doubles from $v/c=-0.0222\pm0.0021$ ($-6700\pm600$\,km\,s$^{-1}$) in slice 1, 
to $v/c=-0.0405\pm0.0022$ ($-12200\pm700$\,km\,s$^{-1}$) in slice 5. The changes in velocity, column and 2--10\,keV luminosity are plotted in Figure~4 (left panels). 
The most pronounced change in the velocity occurs between slice 3 and 4 and coincides with the onset of the strong X-ray flaring in slice 4. 
In contrast, the column density is approximately constant. 

Thus the wind acceleration may be the cause of the 
energy shift of the absorber.  
From a linear fit of velocity versus time across all five slices, the acceleration is then $a=2500\pm400$\,cm\,s$^{-2}$. 
This assumes a constant acceleration, however a more rapid increase in velocity occurs from slice~3 to 4. From comparing the midpoints of these slices, this yields 
$a = 9300\pm2100$\,cm\,s$^{-2}$. Thus the acceleration covers a likely range from $2.5-10.0\times10^{3}$\,cm\,s$^{-2}$, equivalent to $2.5-10{\rm g}$ at the Earth's surface.  

The null hypothesis scenario where the absorption is not variable was also tested. If first the column density is fixed to a constant value across all of the slices, then the fit statistic is very similar ($C/\nu=1804.3/1813$), as expected if the opacity remains similar through the observation. On the other hand, if the velocity is also held constant, then the fit statistic is significantly worse, with $C/\nu=1844.9/1817$. The 
change in the fit statistic is then $\Delta C/\Delta\nu=-40.6/4$, equivalent to $\Delta {\rm AIC}=-32.6$ and a null probability of $P_{\rm AIC}= 8.3\times10^{-8}$. 

As is noted in Section~4.1, slice 5 appears to show a higher energy absorption trough centered just below 8\,keV. Adding in an additional \textsc{xstar} absorber to model this trough results in a much higher 
outflow velocity of $v/c=-0.1336\pm0.008$ (see Table~3). While the additional of this high velocity zone is formally significant, with $P_{\rm AIC}=1.4\times10^{-3}$, we caution that it occurs only in the final lower flux slice. 
As a result of this transient behavior, the imprint of this fast zone will be diluted in the mean spectrum, where it is not observed. 
In contrast, the lower velocity zones appear across the slices and thus are detected in the mean spectrum due to their more persistent nature. 
Note that, for the faster zone, its velocity is consistent with previous measurements of a faster UFO with \xmm\ \citep{PV12}.


\begin{figure*}
\begin{center}
\rotatebox{-90}{\includegraphics[width=7cm]{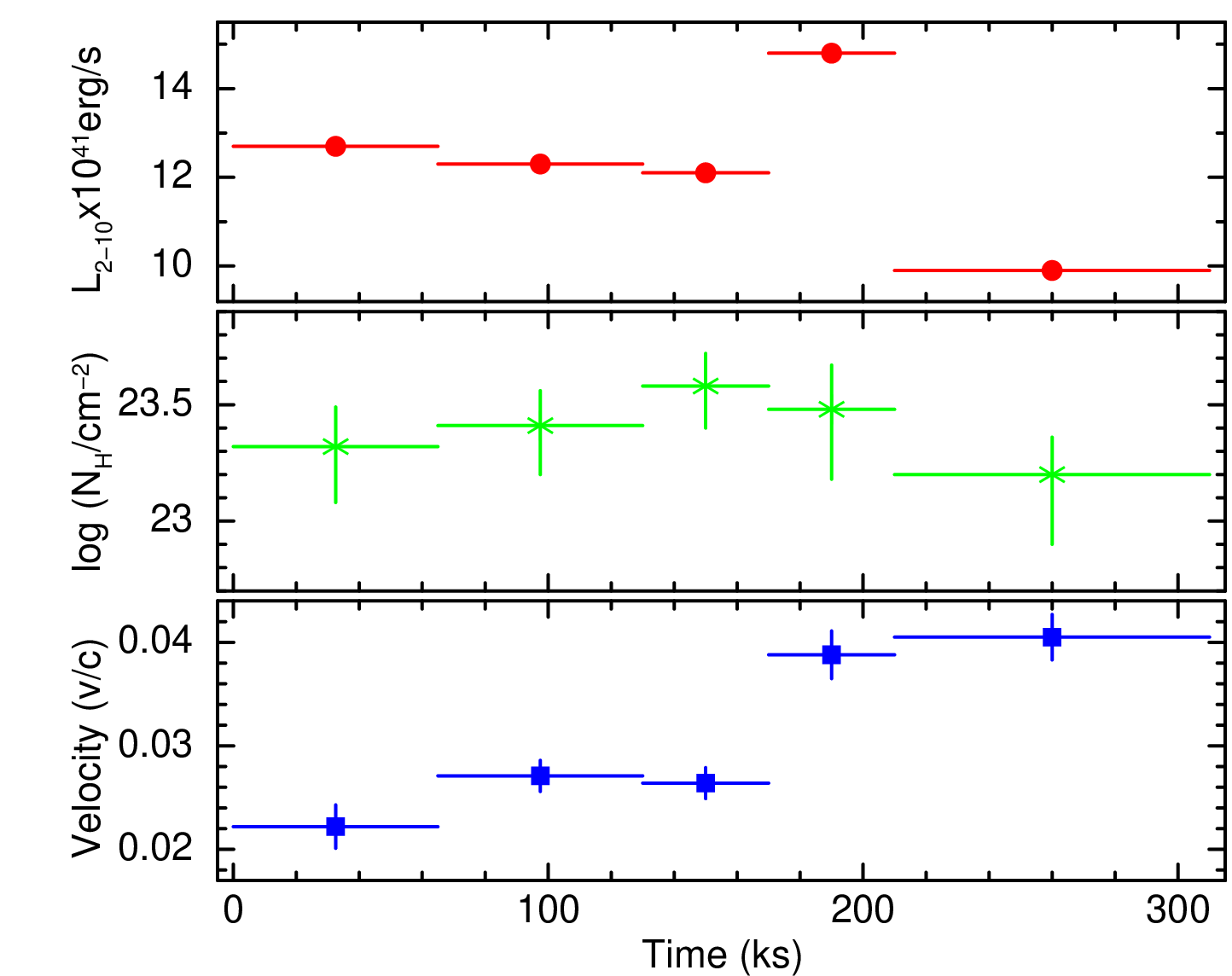}}
\rotatebox{-90}{\includegraphics[width=7cm]{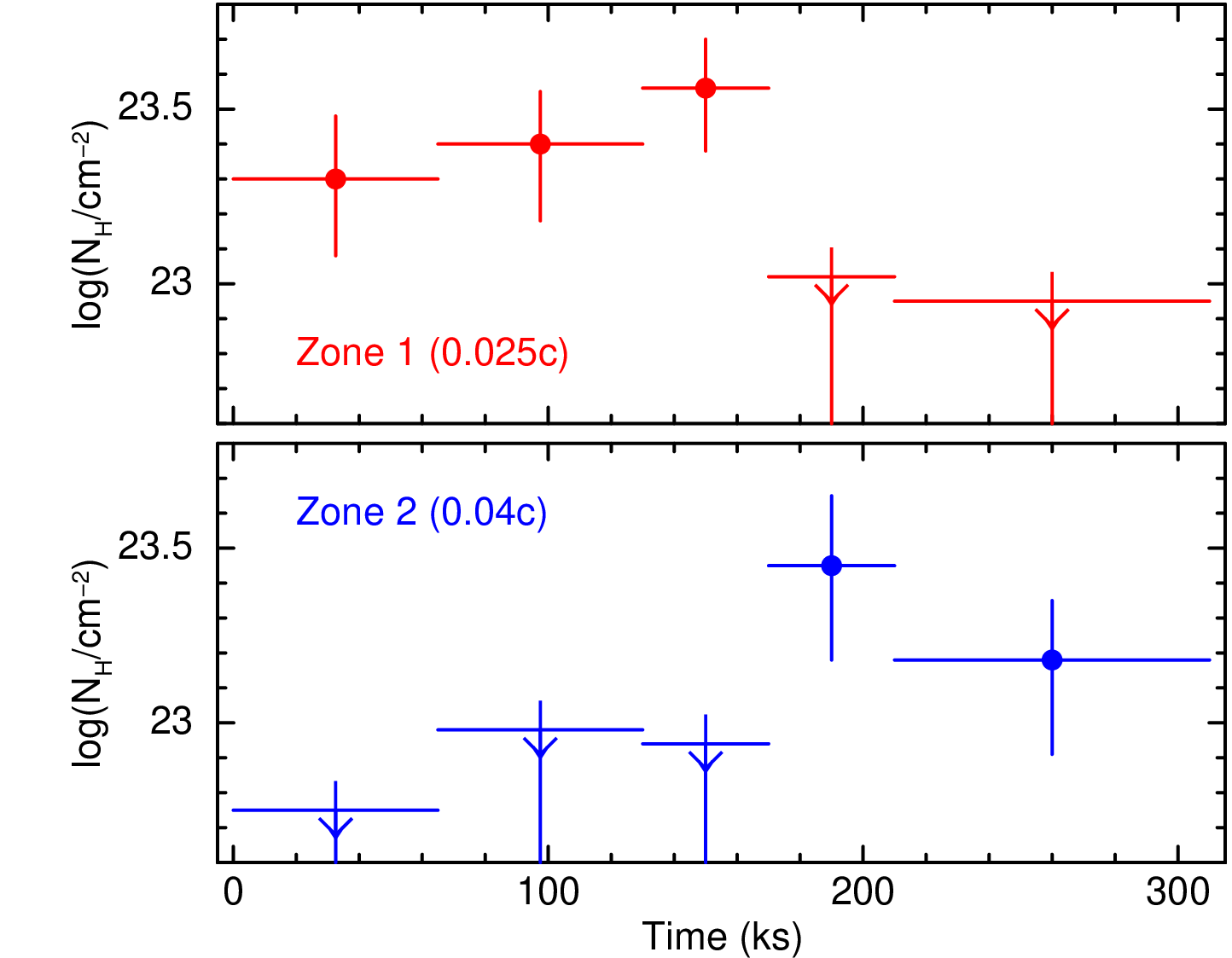}}
\end{center}
\caption{Results from fitting the Resolve slices. The left plot shows the results from fitting a single absorption zone, allowing the outflow velocity to vary. The top panel shows the 2--10\,keV 
luminosity, the middle panel the column density and the lower panel the outflow velocity. The absorber velocity increases through the observation, with a sharp jump occurring between slices 3 and 4. 
The right plot shows the case of fitting two separate absorbers to the slices, where the velocity is restricted to $v/c=-0.025\pm0.005$ and $v/c=-0.040\pm0.005$ for each zone. Here, changes in the column density of the two zones accounts for the variability. 
Thus the slower zone is apparent during the first 3 slices and is not detected in slices 4 and 5. Conversely, the faster absorber is only apparent in slices 4 and 5, with upper-limits only during slices 1--3.}
\label{fig:absorbers}
\end{figure*}

\subsection{Modeling with two transiting absorbers}

Alternatively, instead of one accelerating outflow, the Fe K absorption may arise from two independent absorption zones of different velocities 
which transit across the X-ray source at different times. 
To test this hypothesis, two \textsc{xstar} absorbers were added, allowing the column of each zone to vary across all 5 slices. In this scenario, the absorber velocities were restricted to a narrow range,  
with $v/c=-0.025\pm0.005$ for zone 1 and $v/c=-0.040\pm0.005$ for zone 2, such that the varying profiles cannot be accounted for by large velocity variations.   
An equally good fit was obtained with this model set-up, with $C/\nu=1795.8/1804$. 
The resulting column density variations of the two zones is plotted in Figure~4 (right panels) and are listed in Table~4. 

\begin{deluxetable}{lcc}
\tabletypesize{\small}
\tablecaption{Column Density Variations for Two Transiting Absorbers}
\tablewidth{0pt}
\tablehead{
\colhead{Slice} & \colhead{Zone 1$^a$} & \colhead{Zone $2^b$} \\
\colhead{} & \colhead{$\log (N_{\rm H} / {\rm cm}^{-2})$} & \colhead{$\log (N_{\rm H} / {\rm cm}^{-2})$}
}
\startdata
1 & $23.3^{+0.18}_{-0.22}$ & $<22.85$\\
2 & $23.4^{+0.15}_{-0.22}$ & $<23.08$\\
3 & $23.56^{+0.14}_{-0.18}$ & $<23.04$\\
4 & $<23.12$ & $23.45^{+0.20}_{-0.27}$\\
5 & $<23.04$ & $23.18^{+0.17}_{-0.27}$\\
\enddata
\tablenotetext{a}{Zone 1, $v/c=-0.025\pm0.005$}
\tablenotetext{b}{Zone 2, $v/c=-0.040\pm0.005$}
\label{tab:twoabs}
\end{deluxetable}

In the two zone scenario, the lowest velocity ($-0.025c$) absorber dominates the Fe K band opacity in slices 1--3, while it is not detected in slices 4 and 5. The converse is the case for the faster $-0.04c$ zone, 
which it is not present during slices\,1--3 but is detected in slices 4 and 5. Here the two absorbers are likely to be different outflowing streamlines or clouds, which cross the X-ray source 
during different portions of the observations at an approximately constant velocity but with varying columns. 

Then what is the chance probability that the superposition of the two transiting absorbers can produce the observed velocity increases across the five slices? 
This requires a coincidental timing, whereby the slower absorber disappears exactly as the fast absorber appears.
The statistical probability of this is estimated from a probabilistic argument. 

Consider five slices, whereby either a slow (S, 0.025c) or a fast (F, 0.04c) absorber can be detected with approximately equal likelihood. 
The total number of possible permutations and combinations across these slices are $2^5 = 32$. These are 5S or 5F, 4S and 1F or 1S and 4F (5 combinations each), 3S and 2F (10 combinations) and 2S and 3F (10 combinations). 
By random chance, to match the observed behavior, we would require the slow absorber to transition to a fast absorber across the 5 slices.  These could occur in the order SFFFF, SSFFF, SSSFF (the case observed here) and SSSSF. Thus, four possible combinations are permissible. Overall, the chance probability of reproducing the behavior is then 4/32 or 0.125. For 5 slices, this is not low enough to rule out the transiting case with high certainty. However, a greater number of slices, e.g. from obtaining a longer follow-up observation, would help discriminate between the competing scenarios. Obtaining ten slices would reach the point where by random chance the probability is $<1$\%.

\section{Discussion} \label{sec:discussion}

High resolution X-ray spectroscopy of Fe K outflows in AGN with \xrism\ is revealing a wealth of outflow components in unprecedented detail. The emerging picture is that AGN outflows are structured, clumpy, cover a wide range of velocities and can sometimes be time variable \citep{xrism25,Mehdipour25,Noda25,Mizumoto25,Xiang25,Juranova25,Gu25,Brenneman25,Z26}.  

The \xrism\ Resolve spectrum of NGC\,4051, which occurred in a high flux state, has revealed at least two variable Fe K absorbers with velocities of 0.025c and 0.040c. 
A transient detection of an even faster 0.13c zone appears at the end of the observation (slice~5). 
The main Fe K absorption variability is manifested by an increase in velocity from $0.02-0.04$c over a 300\,ks timescale. 
There are two possible origins for the velocity increase; either, (i) through the separate transits of two absorbers of different velocities or, (ii) through direct acceleration of the outflow in a single streamline. Both scenarios can provide an equally good fit to the data and are discussed below. 

\subsection{Transiting absorber scenario}

In the transiting case, the slower ($0.025c$) absorber crosses the AGN during slices 1--3 (0--170\,ks) and the faster outflow (0.04c) follows during slices 4--5 (170--310\,ks). 
The crossing timescale of the transiting absorbers is $\Delta t \approx 150$\,ks. 
The corresponding absorber size is $\Delta R = v_{\rm trans} \Delta t$, where the transverse velocity can be set to the Keplerian velocity as $v_{\rm trans} = (GM/R)^{1/2}$ and $M$ 
is the black hole mass. Furthermore, the column density, number density $n$ and absorber size are related as $N_{\rm H} = n\Delta R$, while the 
ionization parameter is defined $\xi = L_{\rm ion}/nR^2$, where $L_{\rm ion}$ is the 1--1000\,Rydberg ionizing luminosity. 

Combining these equations leads to an expression for the radial distance of the transiting absorber:- 

\begin{equation}
R^{5/2} = \frac{L_{\rm ion}}{N_{\rm H}\xi} (GM)^{1/2} \Delta t.
\end{equation}

\noindent From the spectral fits (Tables 2 and 3), $N_{\rm H}=2\times 10^{23}$\,cm$^{-2}$, $\log\xi=4.8$ and from the SED, $L_{\rm ion}\approx10^{43}$\,erg\,s$^{-1}$.
For NGC\,4051, the black hole mass is $\approx10^{6}\,{\rm M}_{\odot}$ \citep{Denney09,Fausnaugh17}.
Thus for a transit timescale of $\Delta t = 150$\,ks, then the radial location of the outflowing gas is $R\approx 3\times10^{14}~{\rm cm} \approx10^3~{\rm R}_{\rm s}$ (${\rm R}_{\rm s}$ is the Schwarzschild radius) or equivalently $\sim0.1$\,light-days. The Keplerian transit speed is then $v_{\rm trans} \approx10^4$\,km\,s$^{-1}$.

This distance estimate is an order of magnitude inside the inferred broad line region (BLR) radius in NGC\,4051, of $1.9\pm0.5$ light-days \citep{Denney09} and is well inside its dust sublimation radius of 0.01\,pc 
\citep{Krongold07}. 
The distance is consistent with the time lags of the hard X-ray reprocessor in archival \nustar\ data, of a few thousand seconds \citep{Turner17}.

Thus the Fe K absorbers are likely to be located within the optical BLR and could originate from an accretion disk wind \citep{PK04}. 
At this distance and from the above expressions, then the gas density is $n=2\times10^{9}$\,cm$^{-3}$ and the absorber size-scale is $\Delta R \approx 10^{14}$\,cm, which is of the order of its radial distance. 
At $R\approx 10^{3}~{\rm R}_{\rm s}$, the escape speed is $0.03c$, similar to the 
outflow velocities of the absorbers. In this context, the faster (0.13c) zone may arise from closer in, where its escape radius is at $\approx 50$\,R$_{\rm s}$.

\subsection{Constraints on an accelerating outflow}

As the absorber column density remains approximately constant through the observation (Table~3 and Figure~4), this might suggest a single, accelerating outflow. 
Here, the assumption of a Keplerian transit velocity can be relaxed and the velocity variations instead derive from the in-situ acceleration of the gas. 
However, the absorber size-scale should be similar to (or larger) than the size of the X-ray corona, in order for the absorbing gas to fully cover the continuum emission. 
If the absorber is too compact with a low line-of-sight covering fraction, then the absorption profiles will become too weak to be detectable. 

An estimate of the X-ray coronal size derives from the flux doubling time, which is $\Delta t \sim 1$ ks from the Xtend light-curve in 
Figure~1. From the light crossing time, the coronal size is inferred to be $D_{\rm cor}  = c \Delta t = 3 \times 10^{13}$\,cm (or $D \sim 100 R_{\rm s}$). Assuming the absorber fully covers the continuum source, then the absorber size ${\rm D_{abs} \approx D_{cor}}$. 
The absorber density is then $n=\frac{N_{\rm H}}{{\rm D}_{\rm abs}} \sim 6\times10^{9}~{\rm cm^{-3}}$. The radial distance of the absorber inferred 
from the ionization parameter is then $R_{\rm min} \approx 2\times10^{14}$\,cm. This corresponds to a minimum absorber distance. The maximum distance derives from the 
condition that its size is of the order of its radial distance (i.e. $\Delta R/R \approx 1$), which is then $R_{\rm max} \ls L_{\rm ion} / N_{\rm H} \xi \approx 10^{15}$\,cm \citep{Crenshaw03}. 

The gas acceleration may be caused by radiation pressure. As the Fe K absorber is highly ionized, arising from mainly H-like iron with a negligible force multiplier \citep{Dannen19}, 
then the pressure is imparted through electron scattering. The radiation pressure is simply:-

\begin{equation}
F_{\rm rad} = \frac{L_{\rm bol}}{4 \pi R^2 c}  \sigma_{\rm T} N_{\rm H},
\end{equation}

\noindent where $\sigma_{\rm T}$ is the Thomson cross-section and $L_{\rm bol}$ is the bolometric luminosity. Dividing by mass per unit area ($\mu m_{\rm p} N_{\rm H}$) gives the acceleration:- 

\begin{equation}
a =   \frac{L_{\rm bol}}{4 \pi R^2 c}  \frac {\sigma_{\rm T}}{\mu m_{\rm p}},
\end{equation}

\noindent where $\mu\approx1.4$ is the average atomic mass. From the NGC 4051 SED (see Appendix), $L_{\rm bol} \sim 3\times10^{43}$\,erg\,s$^{-1}$, which is about 25\% of the Eddington luminosity for a 
black hole mass of $M_{\rm BH}=10^{6}$\,M$_{\odot}$. 
Taking the lower of the radial estimates, of $R_{\rm min}=2 \times 10^{14}$\,cm, then yields ${\rm a} \approx 500$\,cm\, s$^{-2}$. This is insufficient by about an order of magnitude to drive the gas in NGC 4051, compared to the observed acceleration in Section~4.2 of 
$a\approx 2.5-10\times10^{3}$\,cm\,s$^{-2}$. At the maximum radius of $R_{\rm max}=10^{15}$\,cm, the predicted acceleration ($a\approx20$\,cm\,s$^{-2}$) is more than an order of magnitude smaller still, incompatible with the observed value. 

To achieve the required acceleration, even at the smaller $R_{\rm min}$ value, the luminosity would have to be at least $\times10$ greater than observed, pushing NGC\,4051 above the Eddington limit. Alternatively the gas could be located further in, closer to 100\,R$_{\rm s}$, but this is problematic as it is well inside the escape radius for gas with a terminal velocity of 0.04c and the absorber size would become too small.

\subsection{An MHD driven wind}
Instead, MHD driving \citep{BP82,Fukumura10} may explain the variable wind in NGC~4051. 
This can be powered by magnetic reconnection events \citep{DiMatteo98}, as is favored to explain the accelerating UFO in NGC\,3783 \citep{Gu25} and 
could coincide with X-ray coronal flares.  
 
In MHD winds, the outflow velocity can be equivalent to the Alf\'{v}en speed ($V_{\rm a}$), which is proportion to the magnetic field strength ($B$) in c.g.s units as:-

\begin{equation}
V_{\rm a} = \sqrt{\frac{B^2}{4\pi\rho}},
\end{equation}

\noindent where $\rho$ is the mass density. From the above estimates for $n$, then $\rho\approx10^{-14}$\,g\,cm$^{-3}$ and hence for $v=0.04$c, 
$B\approx400$\,Gauss. Note for a standard disk prescription \citep{SS73}, then a plausible upper bound is $B\ls10^{3}$\,Gauss, at a distance of $10^{3}\,R_{\rm s}$ and scaled to the black hole mass of NGC 4051. 

The above estimate of $B$ is not strongly dependent upon the absorber radial location. From the ionization parameter, the gas density varies as $n \propto R^{-1/2}$ or equivalently $\rho \propto R^{-1/2}$. 
From equation~4 and for a given Alf\'{v}en speed, then $B \propto \rho^{1/2}$. Thus the dependence between $B$ and $R$ is very weak, varying as $B \propto R^{-1/4}$. For the factor of $\times 5$ uncertainty 
in the radial distance, from $R_{\rm min}$ to $R_{\rm max}$ above, then $B$ only differs by a factor of $\times 1.5$.  

As a consistency check, we can compare the magnetic energy density (equivalent to pressure), $U_{\rm B}$, to the gravitational energy density, $U_{\rm G}$. At a distance of $R\approx10^{3}~{\rm R}_{\rm s}=3\times10^{14}$\,cm, then 
$U_{\rm G}=GM\rho/r \approx 4\times10^3$\,erg\,cm$^{-3}$. For $B=400$\,Gauss, then $U_{\rm B}=B^2/8\pi\approx6\times10^3$\,erg\,cm$^{-3}$ which is of a similar order (or slightly higher). 
In comparison, the radiation pressure is at least an order of magnitude smaller than $U_{\rm B}$, where per unit area $F_{\rm rad}\approx10^{2}$\,erg\,cm$^{-3}$ from equation~2. Hence a Compton (electron scattering) 
driven wind would be unlikely to escape. Thus MHD mechanisms may be more plausible to drive the wind in NGC\,4051. The wind acceleration likely occurs mainly along the line of sight, with relatively small changes in 
column density. 

\subsection{Wind Energetics}

The mass outflow rate can be expressed as:-

\begin{equation}
\dot{M}_{\rm out} = 4\pi b nR^2 v_{\rm out} \mu m_{\rm p},
\end{equation}

\noindent where $b$ is a geometrical factor which is a product of the global covering factor ($f_{\rm cov} = \Omega / 4\pi$) and the volume filling factor $f_{\rm vol}$; e.g. see \citet{Krongold07,Tombesi13,Gofford15}. 
Here $nR^2 = L_{\rm ion} / \xi = 10^{38}$\,cm$^{-1}$ in NGC 4051, $v_{\rm out}=10^4$\,km\,s$^{-1}$ and $\mu=1.4$, which yields $\dot{M}_{\rm out}=0.05b$\,M$_{\odot}$\,yr$^{-1}$. Nominally for $b=1$ and an accretion efficiency of $\eta=0.06$, this is at the 
Eddington accretion rate for NGC\,4051, but is likely lower as $b<1$. 
For example, if $f_{\rm cov}\approx0.4$ from the frequency of occurrence of UFOs \citep{Tombesi10}, while $f_{\rm vol}\approx \Delta R / R \approx 0.3$ from the above constraints on $\Delta R$ and $R$ from the absorber variability, then $\dot{M}_{\rm out}$ is about 10\% of Eddington.

The corresponding kinetic power is $L_{\rm k} = \frac{1}{2} \dot{M}_{\rm out} v_{\rm out}^2 \approx b\times10^{42}$\,erg\,s$^{-1}$, which even for $b=1$ is only 1\% of Eddington in NGC\,4051, with a mass of $10^{6}$\,M$_{\odot}$. Overall the outflow kinetic power in this low luminosity AGN is much lower than those inferred from \xrism\ observations of more powerful, massive quasars, e.g. PDS\,456 \citep{xrism25}, PG\,1211+143, \citep{Mizumoto25}, IRAS\,05189--2524 \citep{Noda25}, which can reach the Eddington limit.

\section{Conclusions} \label{sec:conclusions}

\xrism\ observed the nearby, low mass NLS1, NGC\,4051, at an historically high X-ray flux. The time-averaged Resolve spectrum reveals two blue-shifted absorption troughs in the Fe K band, associated with H-like iron, with outflow velocities of 0.025c and 0.040c. Overall, the wind power is sub-Eddington. However the outflow components vary through the observation, where the slower component is prevalent in the first part of the observation, 
while the higher velocity absorber is present during the latter half. 
The velocity variations can equally be interpreted in terms of two independent transiting absorbers, or through an accelerating absorber. In either case, the increase in velocity coincides with the brightest, flaring portion 
of the observation. Radiation alone, via electron scattering, appears insufficient to accelerate the gas up to 0.04c. Thus MHD mechanisms are favored to drive the wind, which may be connected to increased 
periods of magnetic activity in the accretion disk corona. Given its pronounced variability on short timescales, future, long \xrism\ observations of NGC\,4051 would be ideal for testing the wind variability and how it responds to the continuum. 

\section{Acknowledgements}

JR and VB acknowledge support through NASA grants 80NSSC25K7845 and 80NSSC22K0474. 
TJT acknowledges support from NASA Grant
80NSSC25K0493. This work was also supported by the Japan Society for the Promotion of Science (JSPS) KAKENHI grant number 24K17104 (S.O.).
ATF was supported by an appointment to the NASA Postdoctoral Program at the NASA Goddard Space Flight Center, administered by Oak Ridge Associated Universities under contract with NASA. 
The {\it XRISM} observation was supported by  observations  with {\it NuSTAR} and {\it XMM-Newton}.
The latter is an ESA science mission with instruments and contributions directly funded by ESA Member States and NASA.

\begin{contribution}
JNR was responsible for writing and submitting the manuscript. JNR, SO, TJT, VB, SY, ATF have worked on the data reduction and analysis. 
TJT and SO are the PIs of the \xrism\ observation.
All other authors contributed to the discussion and a review of the manuscript. 
\end{contribution}

\facility{XRISM, XMM-Newton, NuSTAR}

\software{HEASoft/FTOOLS, XMMSAS}

\newpage
\appendix
\vspace*{-0.5cm}

\restartappendixnumbering

\section{SED modelling}

\begin{figure*}
\begin{center}
\rotatebox{-90}{\includegraphics[width=6cm]{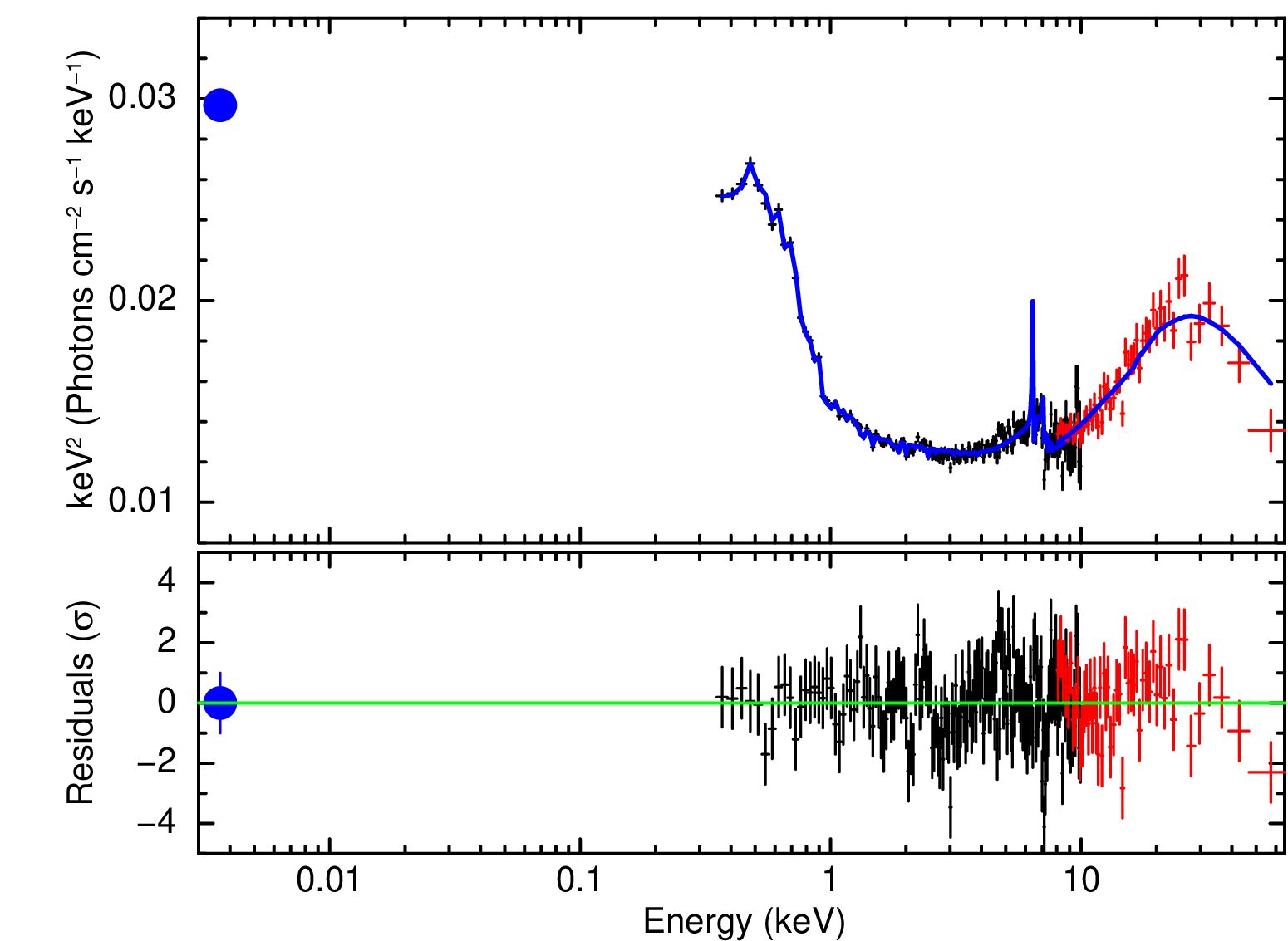}}
\rotatebox{-90}{\includegraphics[width=6cm]{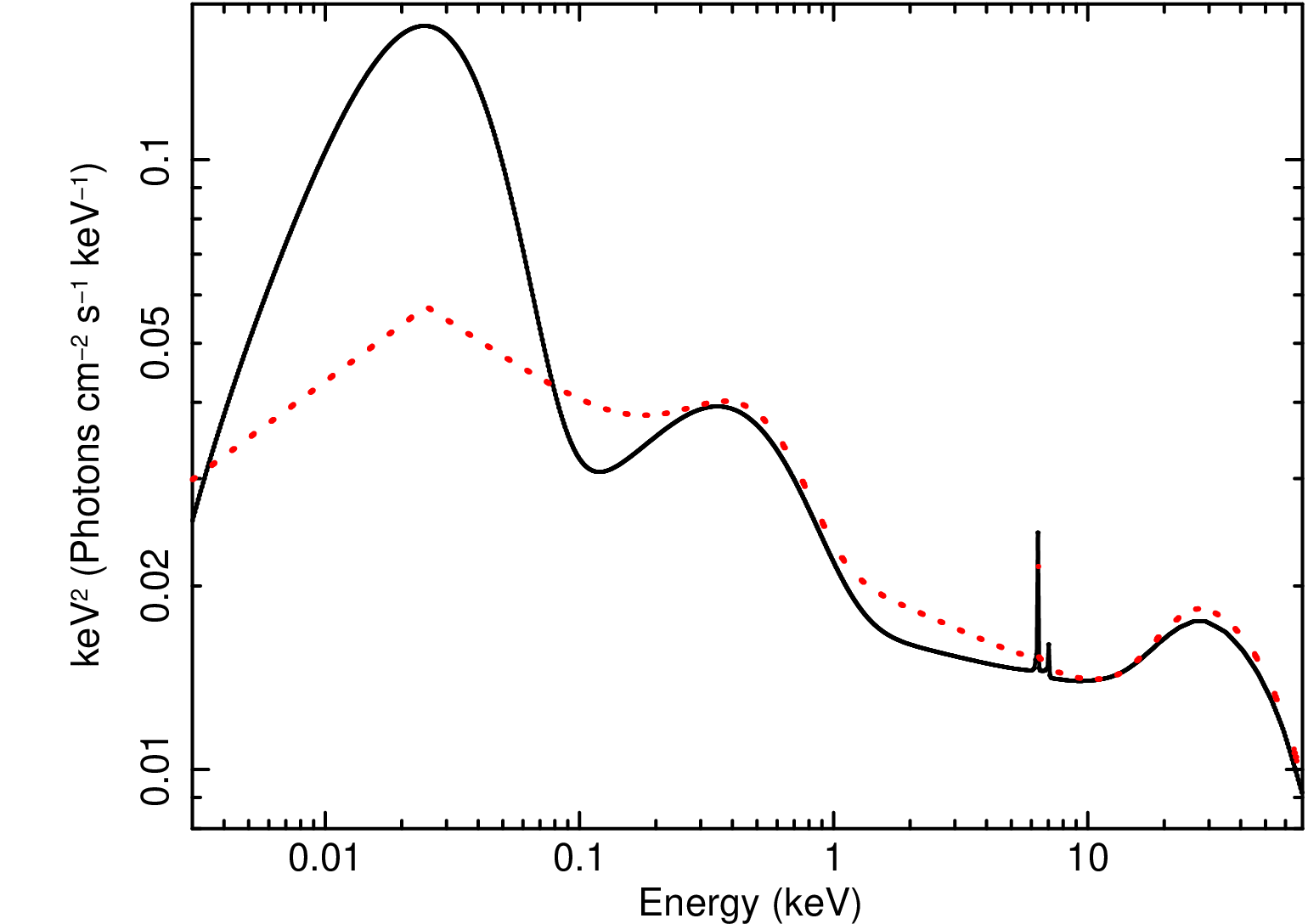}}
\end{center}
\caption{Results from fitting the SED. The left plot shows the fit to the data, where EPIC-pn is black, \nustar\ is red and the OM (U-band) point is a blue circle. The model fitted is the simple 
broken-powerlaw form of the SED, as described in the text. The right panel shows the comparison between the broken-powerlaw SED model (red dotted points), versus a physically motivated model computed 
with \textsc{agnsed} (black). In the latter, the bolometric luminosity is $L_{\rm bol}=3\times10^{43}$\,erg\,s$^{-1}$. 
The models are corrected for reddening and line-of-sight absorption. }
\label{fig:absorbers}
\end{figure*}

The SED was modeled using the simultaneous EPIC-pn, optical monitor (OM), and \nustar\ data. The OM was performed with the U filter and was corrected for Galactic reddening of $E(B-V)=0.013$. 
For a phenomenological description, a broken power-law continuum was used, with a break energy fixed at 25\,eV, with the lower energy portion connecting to the U band; here $\Gamma=1.90\pm0.05$ below 25\,eV and 
$\Gamma=2.16\pm0.02$ above this. The emission and absorption components described in the paper were also included, as well as a soft band warm absorber which will be described in future work. 
A Comptonized disk blackbody function \citep{Titarchuk94} accounts for the soft excess below 2\,keV, of electron temperature $kT=114\pm2$\,eV.  
The results are shown in Figure~A1, where the left panel shows the SED spectral fit and the right panel (red points) the model corrected for line-of-sight absorption. 
The ionizing 1--1000\,Rydberg luminosity of NGC\,4051 from this model is $L_{\rm ion}=1.05\pm0.15\times10^{43}$\,erg\,s$^{-1}$. These values are similar to those found in \cite{Laha14}. 

For comparison, the physically motivated SED model, \textsc{agnsed} within \textsc{xspec} \citep{KD18} is shown in black in the right panel of Figure~A1. 
The model self-consistently predicts the UV bump emission from the disk (Figure~A1, right), as well as the warm and hot Comptonized components; see \cite{KD18} for details. 
For an input black hole mass of $10^6$\,M$_{\odot}$, the accretion rate was determined to be $\log(\dot{M}/\dot{M}_{\rm Edd}) = -0.56\pm0.06$, or an Eddington ratio of 
$L/L_{\rm Edd}=0.27\pm0.03$. The predicted bolometric luminosity is then $L_{\rm bol}=3.4\pm0.4 \times10^{43}$\,ergs\,s$^{-1}$.

\section{Absorption line scans}

\begin{figure*}
\begin{center}
\rotatebox{-90}{\includegraphics[height=5.8cm]{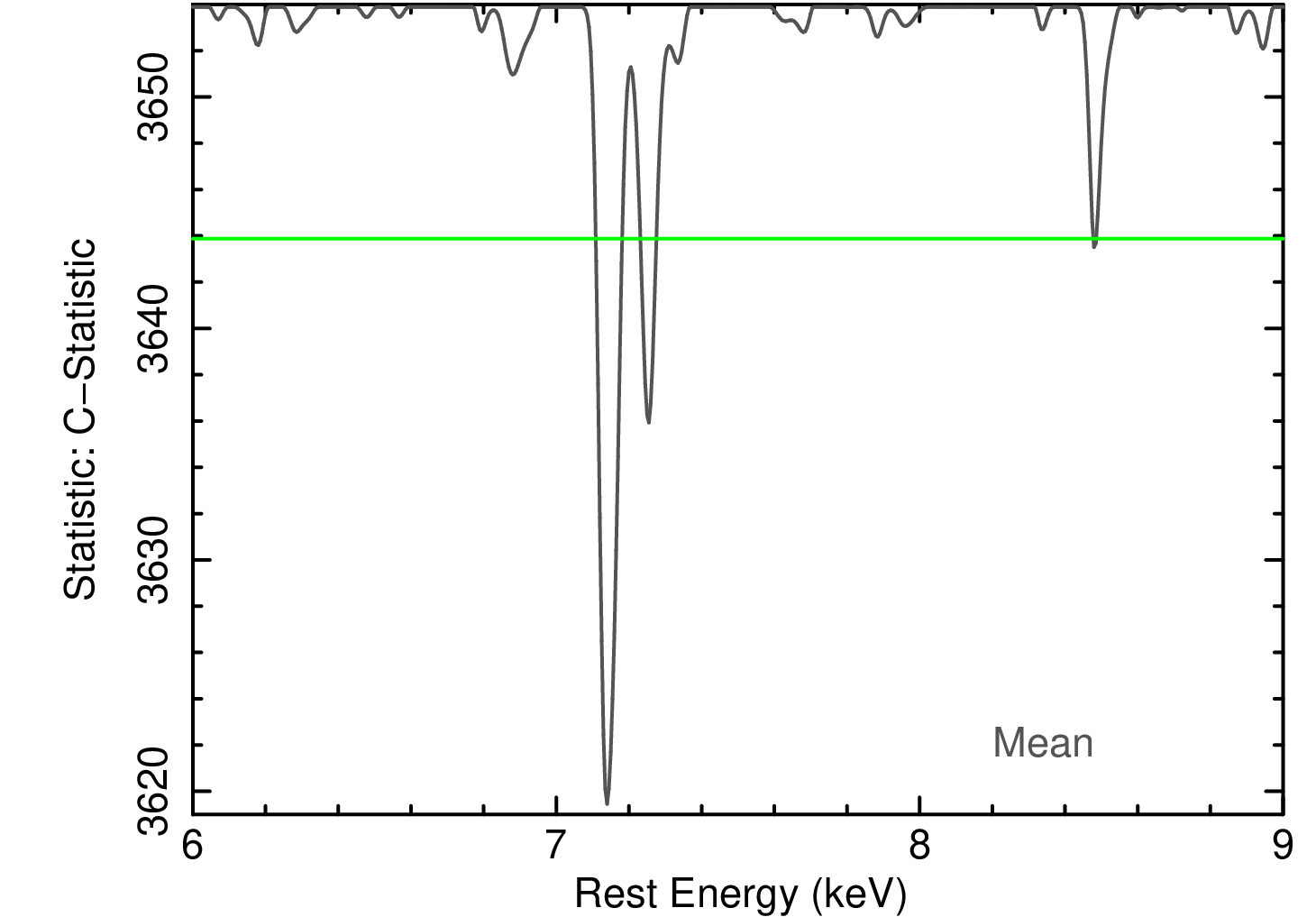}}
\rotatebox{-90}{\includegraphics[height=5.8cm]{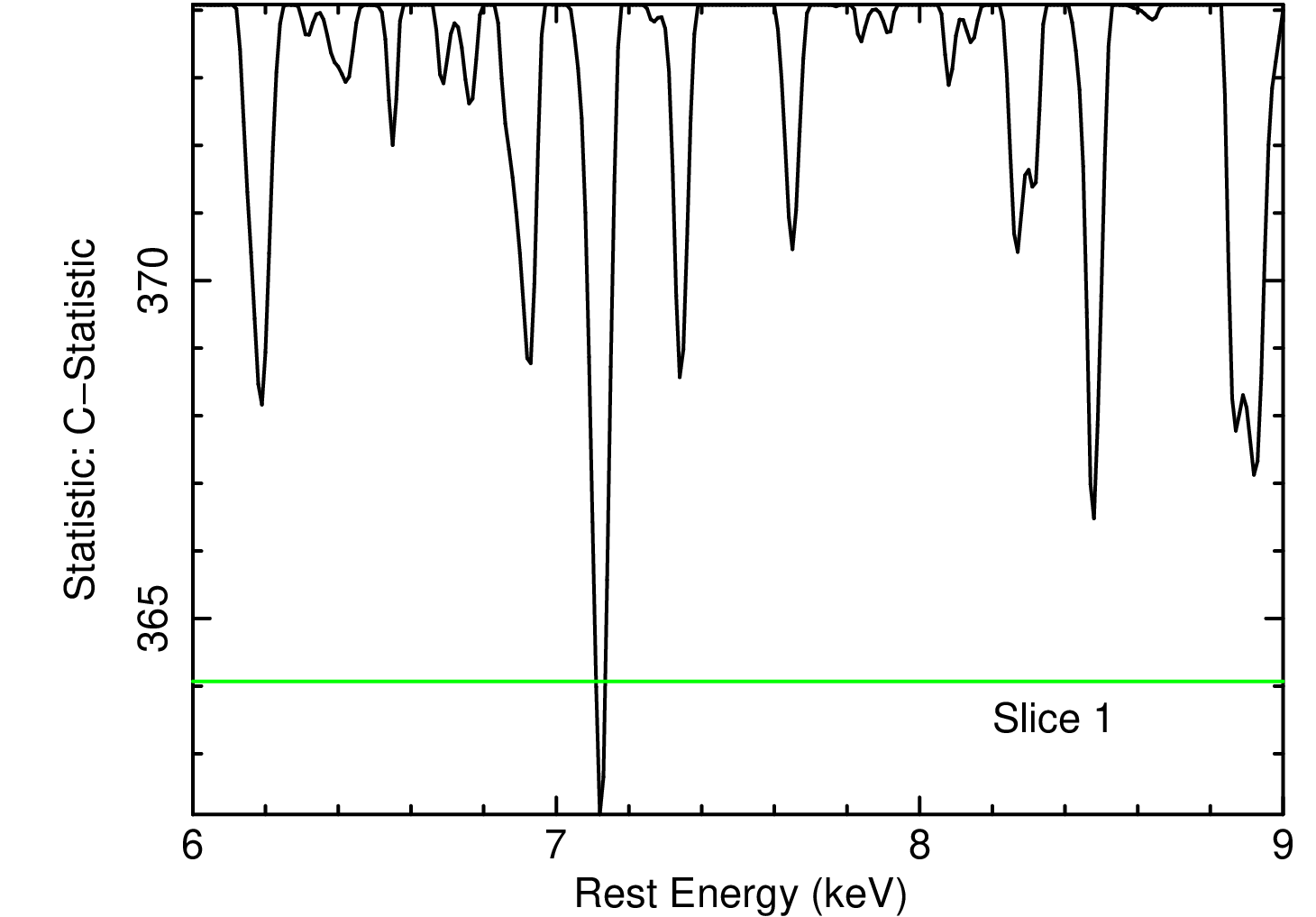}}
\rotatebox{-90}{\includegraphics[height=5.8cm]{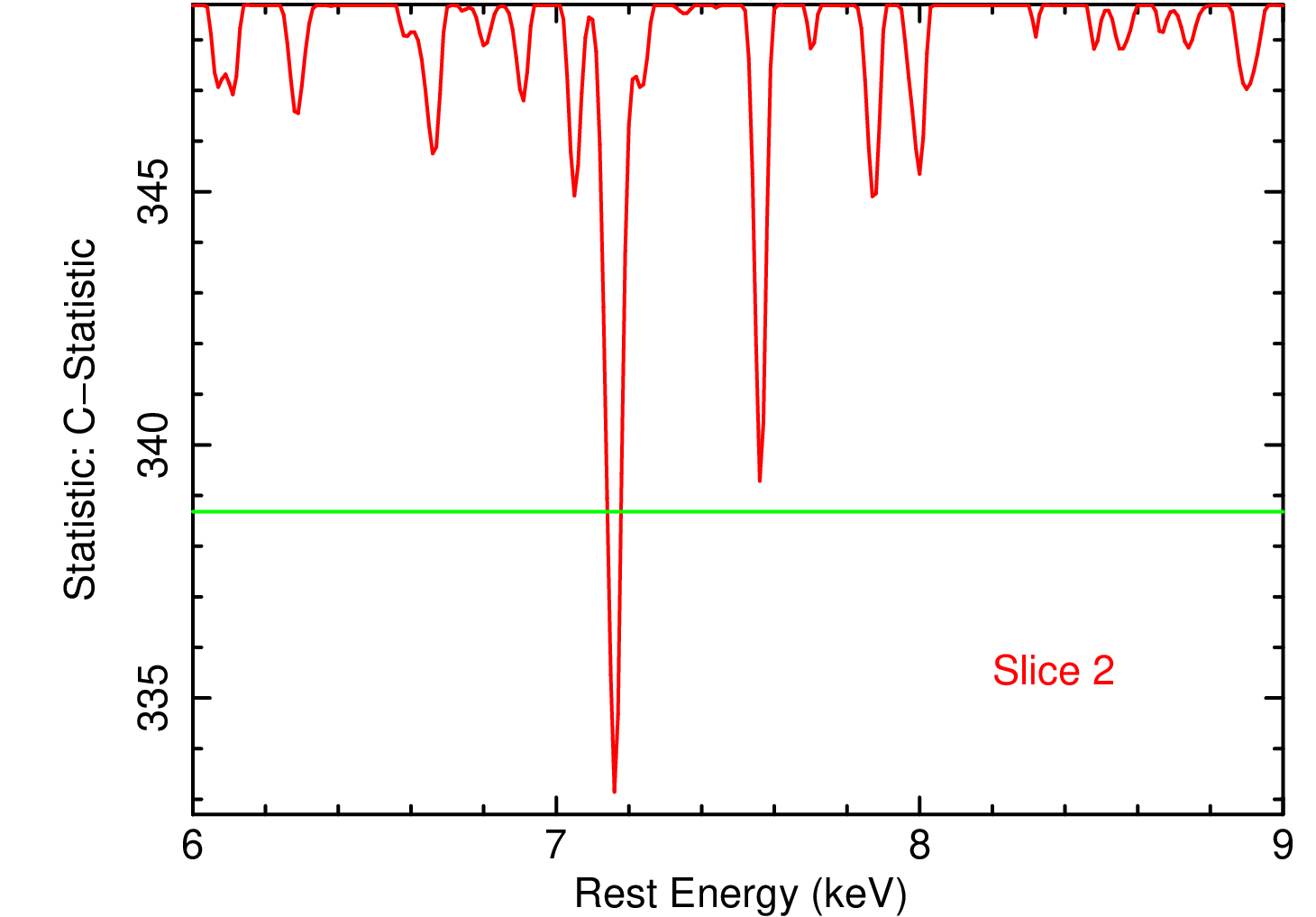}}
\rotatebox{-90}{\includegraphics[height=5.8cm]{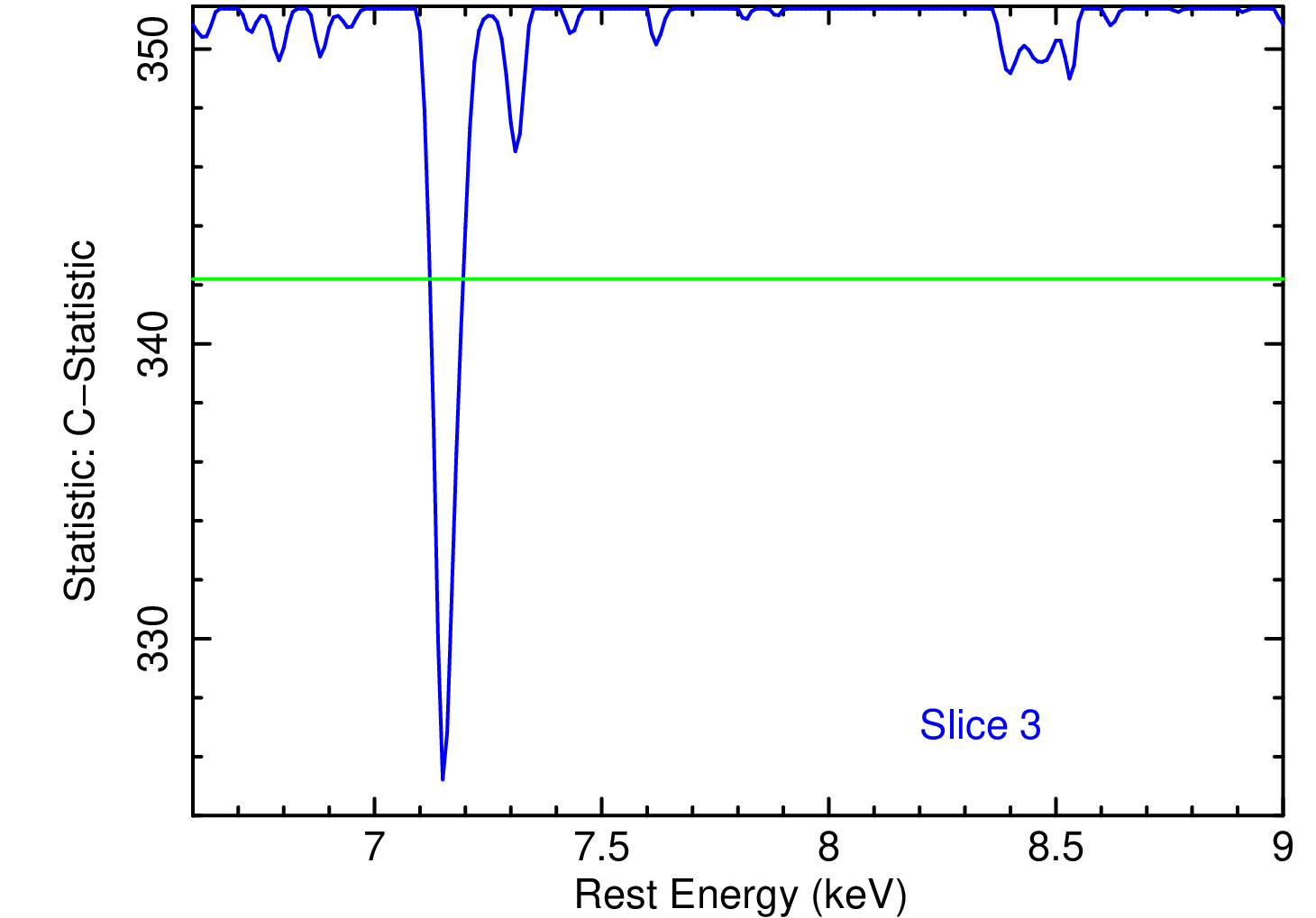}}
\rotatebox{-90}{\includegraphics[height=5.8cm]{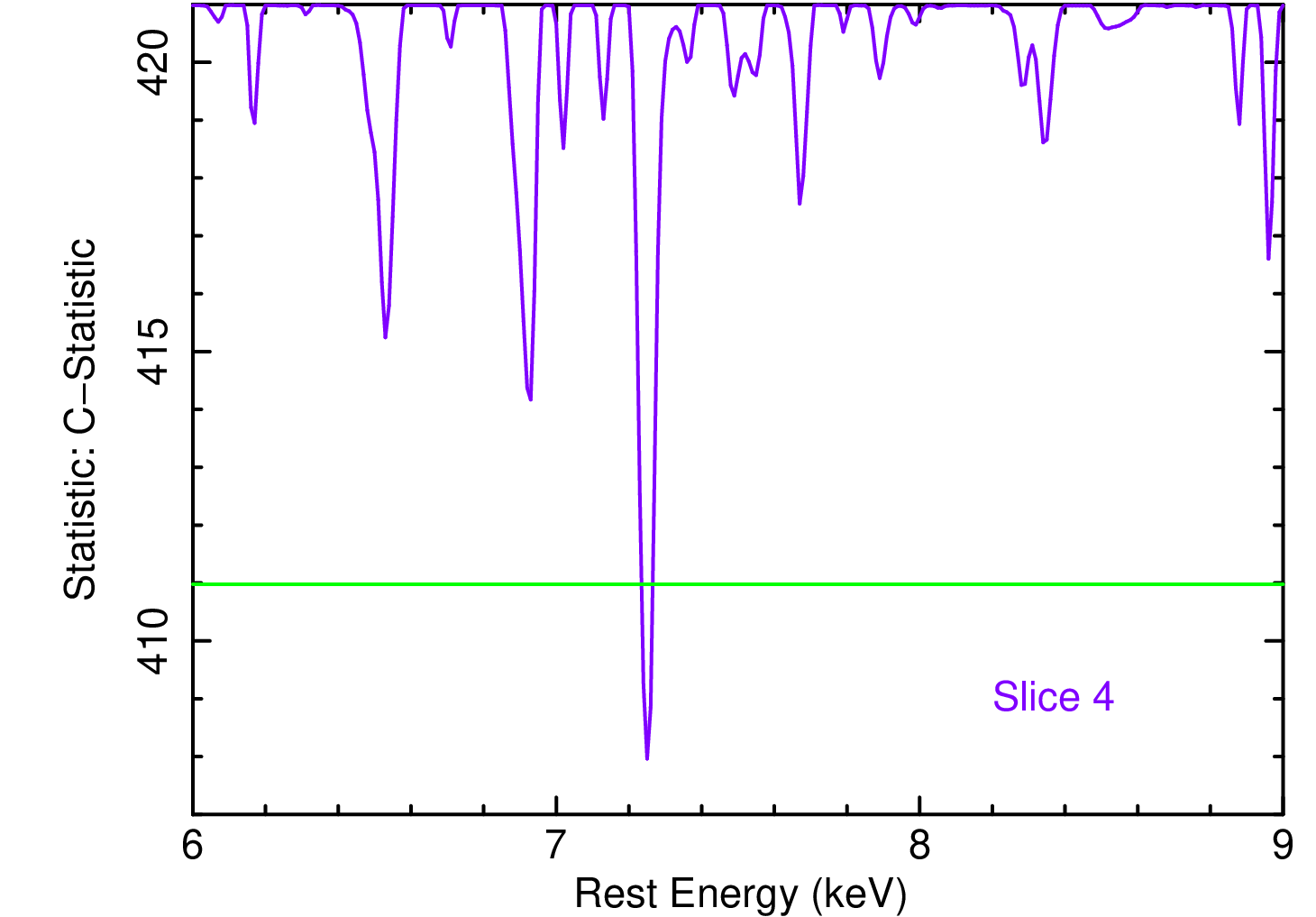}}
\rotatebox{-90}{\includegraphics[height=5.8cm]{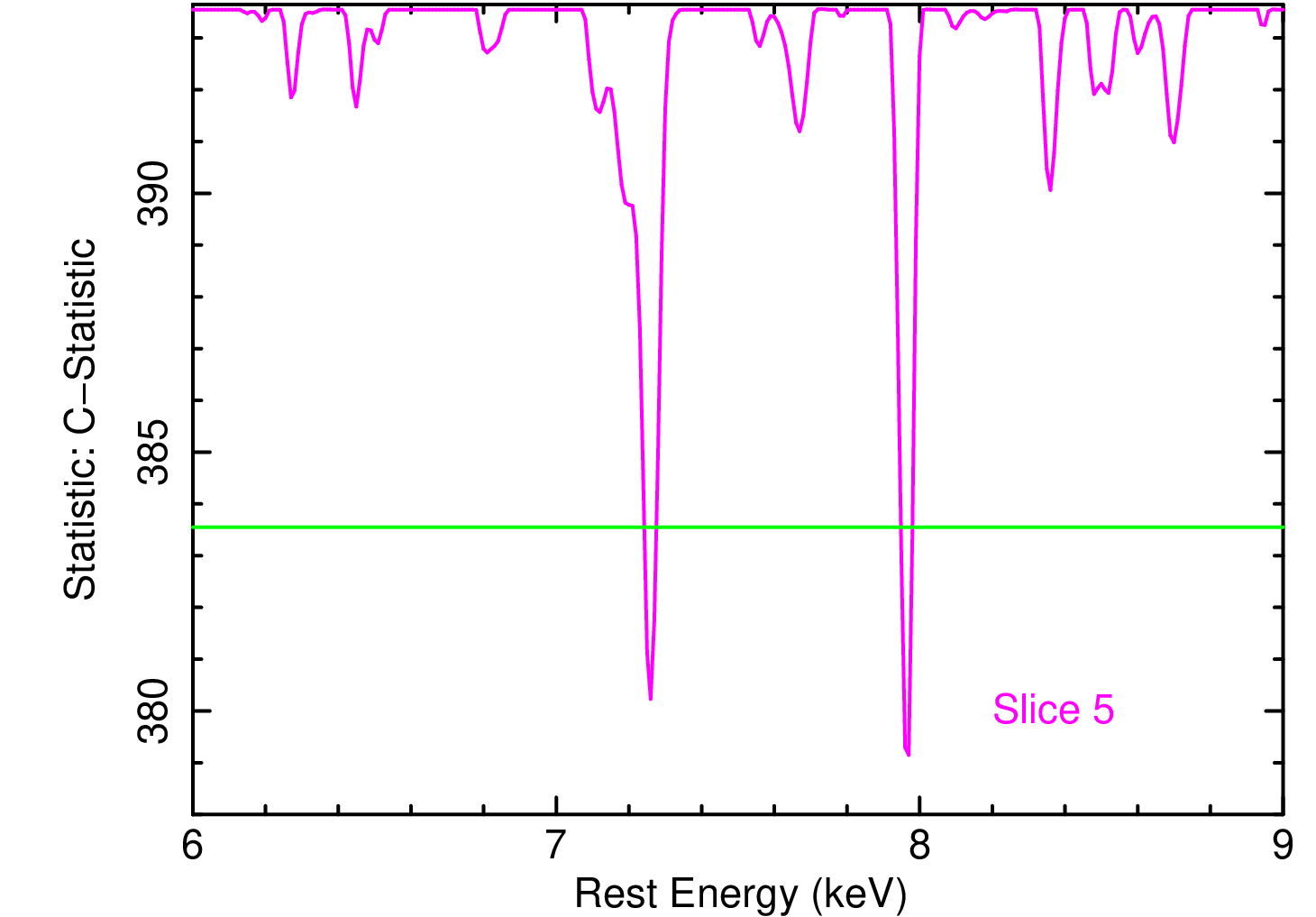}}
\end{center}
\caption{Results of the Gaussian absorption scans lines to the Resolve spectrum, showing the improvement in fit statistic $\Delta C$ against energy. Panel (a) shows the mean spectrum and panels b--f
shows the five time-sliced spectra. The horizontal green line corresponds to a 95\% AIC significance detection threshold (or $\Delta C=10$).} 
\label{fig:appendix}
\end{figure*}

As is described in Section~4.1, scans for Gaussian absorption lines in $\Delta C$ space versus energy, were run on the Resolve spectral slices between 6.0--9.0\,keV. 
The results are shown in Figure~B2 and are listed in Table~3. 
To assess the significance of absorption lines in individual slices, we refer to both the AIC \citep{Akaike74} and also perform blind trial Monte simulations as an independent check. 
For the former, the change in the AIC statistic is $\Delta {\rm AIC} = \Delta C + 2k$, where $k$ is the number of additional fitted parameters and the null probability derives from the maximum likelihood ratio as $P = \exp(\Delta AIC/2)$. 
Thus for the AIC, the 95\% significance threshold corresponds to $\Delta C=-10$ for $\Delta \nu=2$ ($k=2$), with a similar value found from the Monte Carlo estimate described below in Section~C. 
Significant absorption lines are present in all five slices above this minimum threshold, detected between 7.1--7.3\,keV as per the mean spectrum. At higher energies, only the trough near to 8\,keV in slice 5 appears significant at $>95$\% confidence. No other lines are significant. 

\section{Monte Carlo Simulations}

\begin{figure*}
\begin{center}
\vspace{-0.5cm}
\rotatebox{-90}{\includegraphics[height=8.5cm]{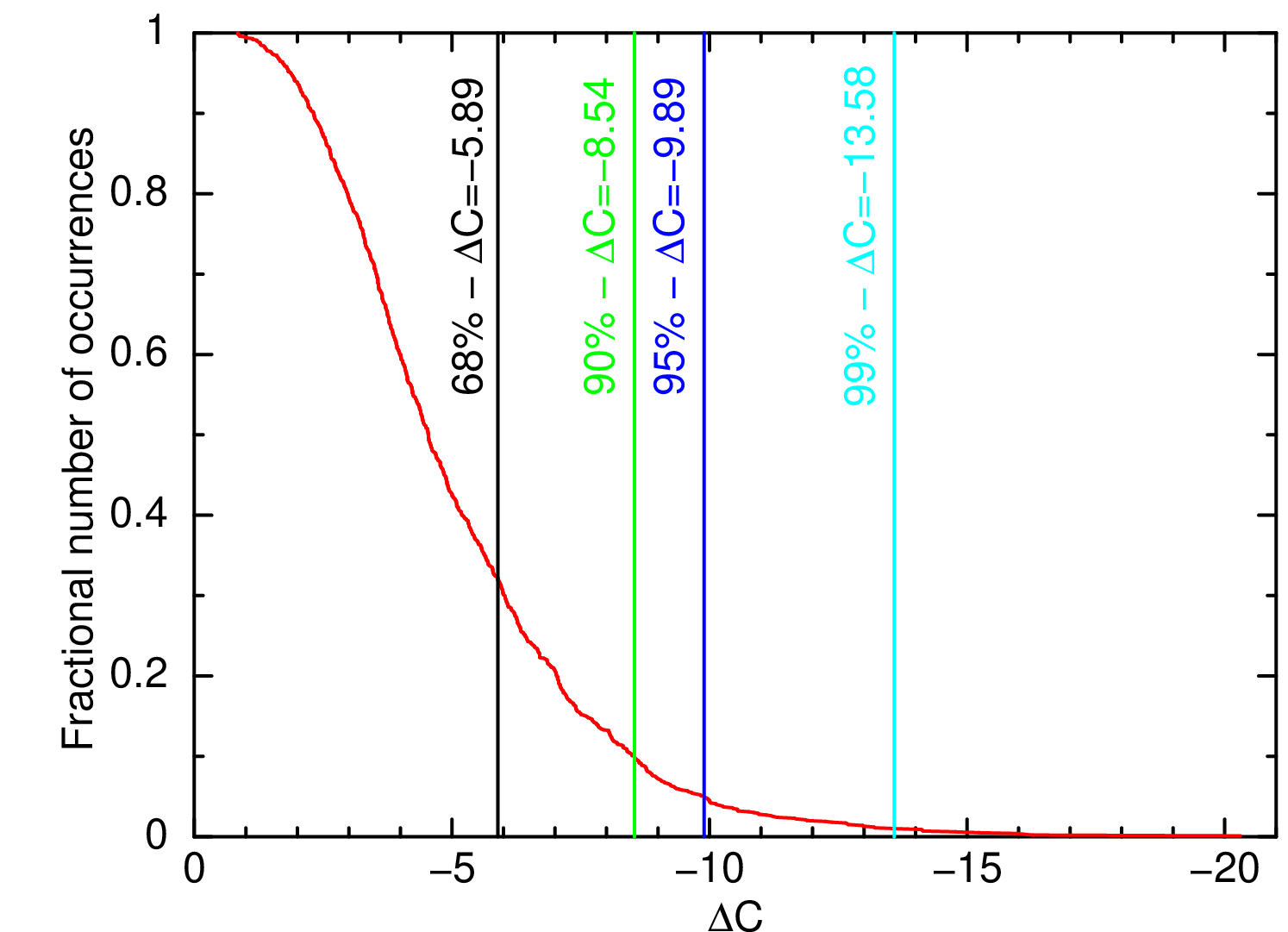}}
\end{center}
\caption{Results from the Monte Carlo simulations for the significance of the absorption lines in the spectral slices, performed over 1000 simulated spectra. The cumulative fractional number of occurrences versus change in C-statistic ($\Delta C$) are plotted.  
The red curve shows the simulated cumulative distribution, while the solid vertical lines show various significance thresholds. For example, the 95\% threshold occurs at $\Delta C=-9.89$, where 5\% of the simulated spectra have the same or higher recorded 
$\Delta C$ values by random chance. The threshold $\Delta C$ values are consistent with those derived from the AIC.}
\label{fig:absorbers}
\end{figure*}

For the Monte Carlo simulations, 1000 random Resolve spectra were generated, adopting the null hypothesis that no absorption lines are present. The continuum model and exposure time from slice~1 was used in the simulations, which is representative of the whole observation. The simulated spectra were fitted, allowing for uncertainties in the power-law continuum and normalization. Each spectrum was fitted with a Gaussian absorption line profile with a free value for the normalization, adopting a line width of $\sigma=20$\,eV. The line energy was stepped in $\Delta E=10$\,eV increments over the 6.9--8.1\,keV band, where the iron K absorption lines might be expected to occur. The maximum value of $\Delta C$ was recorded for each spectrum and the cumulative distribution of the number of false occurrences versus $\Delta C$ was calculated. 
This is plotted in Figure~C3, which shows the corresponding 68\%, 90\%, 95\% and 99\% confidence levels from the simulation. The equivalent values of $\Delta C$ are also noted in Figure~C3; e.g. 95\% of the simulations correspond to a value of 
$\Delta C=-9.89$ or below, while the 99\% threshold corresponds to $\Delta C =-13.58$. These closely correspond to the equivalent thresholds from the AIC, which are $\Delta C=-10.0$ (95\%) and $\Delta C=-13.2$ (99\%). 
Thus the Monte Carlo estimate confirms the significance levels calculated from the AIC. 


\end{document}